**Sculpting 2D Crystals via Membrane Contractions before and during Solidification**

Hao Wan,[1] Geunwoong Jeon,[2] Gregory M. Grason,[1] Maria M. Santore[1,*]

* corresponding author: santore@umass.edu Department of Polymer Science and Engineering, University of Massachusetts 120 Governors Drive Amherst, MA 01003, USA

1. Department of Polymer Science and Engineering, University of Massachusetts
120 Governors Drive Amherst, MA 01003, USA

2. Department of Physics, University of Massachusetts, 710 N. Pleasant St. Amherst, MA 01003, USA

**Sculpting 2D Crystals via Membrane Contractions before and during Solidification**

Hao Wan,[1] Geunwoong Jeon,[2] Gregory M. Grason,[1] Maria M. Santore[1,*]

* corresponding author: santore@umass.edu Department of Polymer Science and Engineering, University of Massachusetts 120 Governors Drive Amherst, MA 01003, USA

1 Department of Polymer Science and Engineering, University of Massachusetts 120 Governors Drive Amherst, MA 01003, USA

2. Department of Physics, University of Massachusetts , 710 N. Pleasant St.  Amherst, MA 01003, USA

**Abstract**
When phospholipids crystallize within the otherwise fluid membranes of giant unilamellar vesicles, the resulting molecularly-thin "2D" solids exhibit great variety in their morphology evolution.  For instance within membranes containing moderate amounts of the crystallizing component, crystals grow with a fixed morphology depending on vesicle size. Conversely for membranes containing large amounts of the crystallizing species, we find small compact crystals on vesicles of all sizes.  However on large vesicles, growing crystals sprout flower petals that lengthen progressively.  These behaviors result from two combined mechanisms:  First, like other 2D solids, the shear rigidity of phospholipid crystals renders them intolerant to morphologies with non-zero Gaussian curvature. As a result and especially at elevated membrane tension, the cost of bending elasticity is reduced, at the expense of line energy, by the formation of flowers as opposed to compact crystals. Second, the composition-dependent tension rise during cooling relaxes via water permeation of the membrane with a time constant scaling as $R^2$.  The amount of crystal formed for a small decrease in temperature determines this composition-dependent increase in stress from thermal contractions versus solidification. Surface Evolver computations motivated using the predicted tension evolution to develop a processing space that maps to experimental observations for initial and growing crystal morphology.  Important variable groups are identified, including a scaled ratio of bending to line energy, a vesicle size-independent group for membrane contractions, and a time constant for stress relaxation. Though processing stresses ultimately relax, the crystal morphology persists well beyond the processing window.

**Significance statement**
This work broadens our understanding of the impact of tension in biomembranes and lamellae, especially on the morphology of membrane phases having shear rigidity, including rigid but molecularly disordered membrane rafts. Additionally, since lamellar phospholipid crystals share shear rigidity and phase transition fundamentals with other 2D materials, from atomically thin graphene to 2D colloidal crystals, the strategy to control morphology through the stresses arising from interactions with a template or support, can be quantitatively generalized to 2D molecular and colloidal systems, requiring only knowledge of their physical properties, including their thermal expansivity, area change on crystallization, moduli, and stress relaxation kinetics. This approach could ultimately enable new methods to produce 2D materials with complex, bespoke morphologies.

**INTRODUCTION**

In studies of biomembranes, the crystallization of high-melting temperature species within mixed phospholipid bilayers and vesicles has revealed an array of morphologies that, if better understood, could more broadly advance the field of ultrathin or "2D" materials.[1-6] Phospholipid crystals, with their ~4 nm thicknesses, constitute a mid-range thickness quasi-2D material, exemplifying a molecular scale that is intermediate within a spectrum from atomically thin layers such as graphene, to monolayer colloidal crystals having thicknesses from ~50 -1000 nm. 2D solids generally prefer to be flat and for phospholipids, their finite thickness in combination with a preferred equal lipid density in both leaflets of the crystal leads to bending resistance.[7-11] Besides order, critical distinguishing features of 2D solids exemplified by phospholipid crystals include shear rigidity[12] and incompatibility with non-zero Gaussian curvature,[13-17] the imposition of which results in defects or cracks.[13, 18-23] In the case of graphene or monolayer crystals such as $WS_2$ and also with supported phospholipid bilayers,[20-22, 24] curvature arises from interactions with roughness and raised substrate topography; in interfacially-assembling colloidal crystals curvature can arise from surface tension of droplet templates;[18] and in free standing phospholipid bilayers and vesicles curvature may result from forces on elastic lamellae.[3, 25, 26] Studies in all three scenarios establish the how avoidance of non-zero Gaussian curvature alters the morphology of growing 2D crystals, even when curvature seems small, for instance corresponding to droplets and vesicles with diameters exceeding 10 microns.[18, 27] Simulations have shown how solid mechanics and interactions between crystallizing units, on different length scales from atoms to colloids, dictate morphology.[17, 20, 28, 29]

Compounding curvature-induced elastic distortions, thermal stresses are generally encountered in the fabrication of thin films and 2D materials. For instance, mismatch of the coefficient of thermal expansion between a graphene film and its substrate can produce wrinkles upon cooling,[30] while stresses develop and become trapped when polymer films are cooled.[31, 32] Thermal stresses are typically overlooked for phospholipid lamellae; however as a result of the substantial alkane portion of these molecules, phospholipid bilayers possess coefficients of thermal expansion in the range 1-5 x $10^{-3}$ °C$^{-1}$.[12] Additionally, crystallization itself causes a substantial reduction in bilayer area, on the order of 15-20%.[12] Thus, when phospholipid crystals nucleate and grow within fluid phospholipid membranes in response to a thermal quench, there is potential for membrane contractions to generate significant stress on the timescale of crystallization. Such stresses occur in supported bilayers[33-37] and, in vesicles they are coupled to curvature.[16, 26, 38] For instance, the vesicle size dependence of crystal morphology in otherwise fluid vesicle membranes includes a compact morphology on small vesicles and shapes with protrusions on large vesicles, a trend opposing that seen in colloidal crystallization on droplets.[18] This distinction highlights the difference between the stresses felt by phospholipid crystals in elastic vesicle membranes versus the surface tension experienced by colloidal crystals on droplets. Indeed the presence of an elastic fluid membrane (termed the $L_\alpha$ phase), which integrates solid phospholipid crystals, in Figure 1A, provides a handle for the control of crystal morphology.

It is puzzling how crystals of different shapes and character can be produced at different compositional points within the same region of a phase diagram, impinging on the same solid phase (the same polymorph). A vexing example probed in the current study, in Figure 1B, is the phospholipid crystallization within membranes of different compositions on the same ultimate room temperature tie line. Since the ends of a tie line describe the fluid and solid membrane phases and compositions, one would expect the same solid composition and morphology and not the variety seen in experiments which are composition- and vesicle size-dependent.

The current work establishes how membrane composition impacts phospholipid crystal growth in otherwise fluid vesicle membranes, first addressing the range of morphologies observed for small crystals at shallow quenches inside the two phase envelope, and then tracking morphology along cooling trajectories ending at different compositional points on the room temperature tie line. The different compositions correspond to different proportions of the two phospholipids: 1,2-dipalmitoyl-sn-glycero-3-phosphocholine (DPPC, $T_m$ = 41ºC) the crystal former, and 1,2-dioleoyl-sn-glycero-3-phosphocholine (DOPC) which remains in the mixed $L_\alpha$ fluid membrane. We focus, where possible, on vesicles having only one crystal each, probing morphological evolution during growth, rather than impact of composition on nucleation density, which was studied previously.[39, 40]

Here we build on the mechanism of how shear rigidity, common to all 2D solids, is critical to the selection of crystal shape: The intolerance of 2D crystals to Gaussian curvature sets up a competition between elastic and line energy that selects between compact and extension-containing crystals. The elastic energy of phospholipid vesicles, however, depends on their inflation and therefore membrane tension. Thus by manipulating tension via composition and process history, we ultimately design crystals of select morphology. Specifically for the same initial and final processing temperatures, and same ultimately crystal compositions, in some parts of the phase diagram (ie. at some compositions), large thermal contractions elevate tension in the one phase region while at other compositions stress from the crystallization itself dominates. Stress relaxation, a result of water diffusion across the membrane, proceeds at a vesicle-size dependent rate, so that vesicle size becomes an additional variable to tune crystal morphology.

Experiments traversing different paths through the temperature-composition phase diagram, first at shallow quenches and then separately approaching the room temperature tie line, reveal the shapes of small crystals when they are first visible, and their growth trajectories during cooling. Addressing the morphology of small crystals, a finite-element treatment examines the roles of vesicle inflation, membrane tension, and curvature on the equilibrium solid shape. An engineering model demonstrates how paths through composition and temperature space produce different tension histories that favor different crystal shapes. This translates to an influence of the shapes of phase boundaries on crystal growth as cooling programs are imposed. An interesting consequence is that phase boundaries distant from the actual trajectory exert influence on tension and therefore morphology. This mechanism leads to a powerful, potentially scalable strategy to manipulate crystal morphology, exploiting system-specific features of the phase diagram itself. Indeed we identify key terms: a dimensionless group combining vesicle size, bending energy, and line energy; a composition-dependent term weighing the relative importance of thermal- and solidification-driven membrane contractions; and a characteristic relaxation time that imparts an $R^2$ dependence on vesicle radius. The variable contributions of thermal and crystallization contraction-driven stresses, developed here for free-standing vesicle membranes also apply in other scenarios, for instance supported membranes undergoing a phase change. More broadly, the growth of stress through thermal and phase-change driven contractions, and the impact of that stress on the morphology of 2D crystals whose shapes are selected by an intolerance to Gaussian curvature, allows the mechanisms invoked here to be applied broadly to control morphologies in advanced 2D materials.

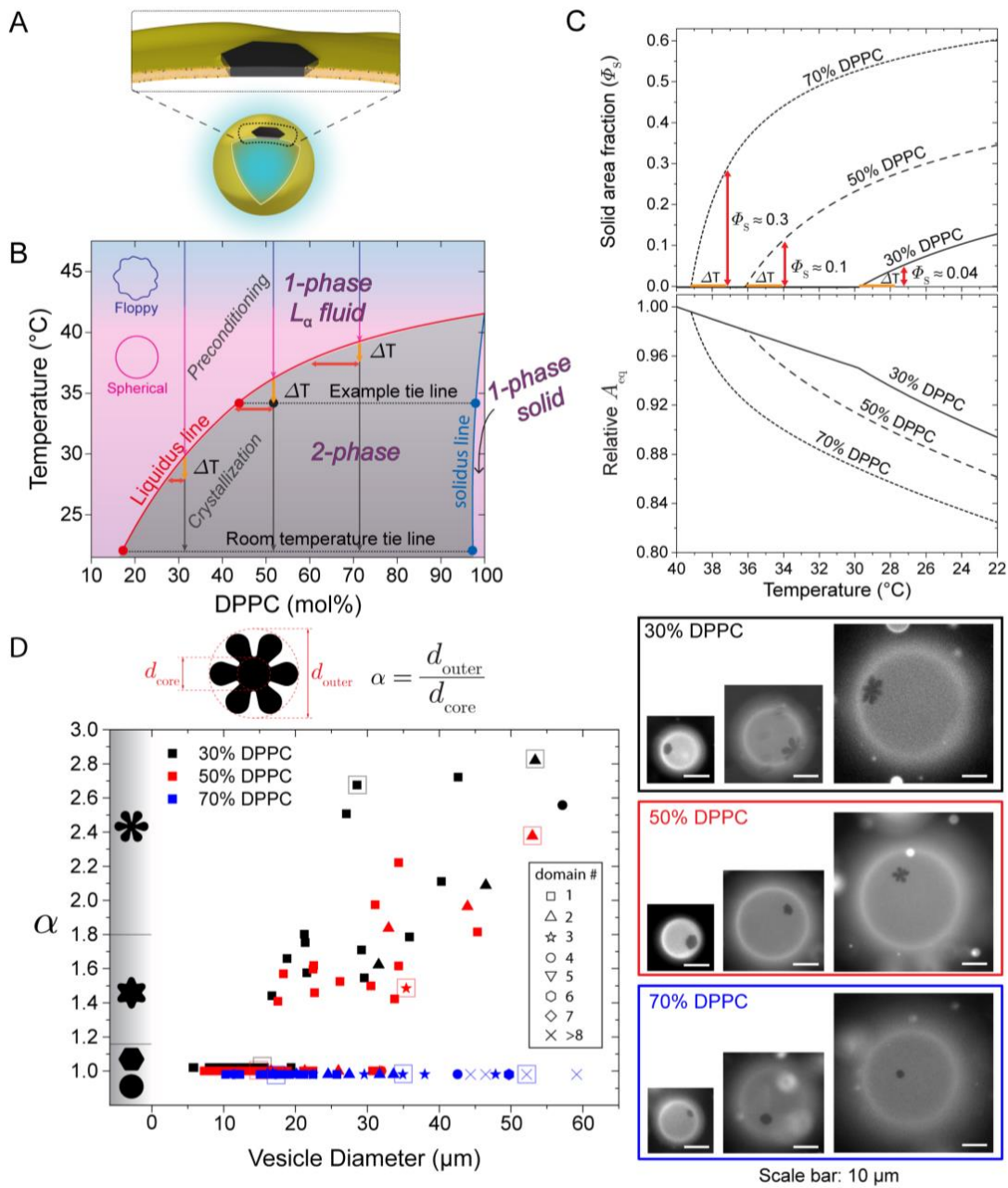

**Figure 1.** A). Schematic of a phospholipid crystal integrated within a fluid phospholipid bilayer vesicle membrane. B) DOPC-DPPC phase diagram, showing 3 cooling trajectories (vertical lines from blue to red to black on each) for different overall membrane compositions, example tie lines, and illustrating, for example quenches of $\Delta T = 2°C$, the equilibrium compositions of the fluid membrane that are read off the liquidus line. C) Top- Equilibrium solid area fractions in the membrane as a function of temperature, calculated from the phase diagram in (B) for the three trajectories highted. Bottom – For same trajectories, the calculated total equilibrium membrane areas relative to that at 40°C. D) Shapes of crystals initially visible on vesicles of different diameters and overall compositions after cooling along the three trajectories in (B). Data are mostly for one crystal / vesicle, with multiple crystals / vesicle indicated by data symbols. Data sets for alpha = 1 are shifted vertically very slightly to enable viewing of the many points that gave these nearly circular crystals. Boxes around 3 datapoints at each composition correspond to the vesicles imaged to the right of the plot. Close-ups of the crystals are found in the SI. Most crystals in D comprise no more 1-2 areal percent of their vesicle, and never more than 4% total when there are multiple crystals with 70% DPPC.

**MATERIALS AND METHODS**

*Experimental System*.  The two phospholipids, 1,2-dipalmitoyl-sn-glycero-3-phosphocholine (DPPC) and 1,2-dioleoyl-sn-glycero-3-phosphocholine (DOPC), along with the tracer lipid 1,2- dioleoyl-sn-glycero-3-phosphoethanolamine-N-(lissamine rhodamine B sulfonyl) (ammonium salt) (Rh-DOPE), were purchased from Avanti Polar Lipids (Alabaster, AL).  Chloroform solutions, with a total of ~5 mg/ml lipid and nominal proportions of DOPC and DPPC by mass were prepared, with addition of no more than 0.1 mol% fluorescent tracer lipid.  These were placed dropwise on the platinum wires of the electroformer, which was then sealed.  A 10 mM sucrose solution preheated to 60°C was then introduced. The electroformer placed on an insulated hot plate, the temperature brought to 55-70°C, and current of 3V and 10 Hz applied for one hour.  The elevated temperature during electroformation ensured that the lipid film and vesicles were in the one phase region of the phase diagram, producing vesicles of uniform targeted composition.  After electroforming vesicles were harvested in a syringe and stored, for up to 3 days, at room temperature.

Crystallization was studied by diluting the vesicle suspension 10-fold in DI water and annealing an aliquot in a custom built temperature control chamber that was attached to a circulating water bath via quick connect fittings. The chamber itself was made of two coverslips (22 x 50 x 1 mm) that were separated by a Parafilm gasket, melted and cooled into place.  The chamber was first heated to 55°C to melt crystals formed during storage and produce uniform membranes.  Then a cooling program was imposed on the chamber and vesicles were observed using a 40x objective on a Nikon Diaphot 300 epifluorescent microscope equipped with a rhodamine cube and a panda-Cam camera.  In some experiments continuous cooling at ~0.3°C/minute to room temperature was conducted and in other cases the cooling was halted just inside the two phase region, within about a degree. Linear cooling was achieved for much of the thermal history, however, approaching room temperature or a targeted intermediate fixed temperature exhibited a modest tail, a reduction in cooling rate towards zero cooling.

Images were captured in two ways. In some instances a single vesicle was tracked throughout the cooling processes, from the moment it was first possible to focus on a crystal, which appeared as a dark spot on a fluorescent vesicle, to completion of the run or through the holding period. In other cases the stage was translated to obtain images of multiple vesicles during cooling.

*Surface Evolver Modeling.* Our theoretical models of fluid-solid composite vesicles were considered as closed membranes including solid potions in the surfaces of fixed solid area $A_{\text{solid}}$, fluid area $A_{\text{fluid}}$, and enclosed volume $V$ where such quantities are set to match desired solid fraction $\phi_s = A_{\text{solid}}/(A_{\text{solid}} + A_{\text{fluid}})$, and reduced volume $\bar{v} = \sqrt{36\pi} V/(A_{\text{solid}} + A_{\text{fluid}})^{3/2}$. We conducted Surface Evolver simulations for three $\phi_s$ values; 0.02, 0.04, 0.08 in the $\bar{v}$ range of 0.98-1.00.

We considered hexagonally symmetric reference solid domain shapes defined as interior to the following radial function

$$r(\theta) = r_0 + \frac{1}{2} a r_0 \cos(6\theta) - \frac{1}{10} a r_0 \cos(12\theta) \qquad (M1)$$

to mimic the experimentally observed solid domain shapes where $a$ was varied to set overall diameter-to-core ratio $\alpha = \frac{5+2a}{5-3a}$, and $r_0$ was set to adjust $A_{\text{solid}}$. We performed modeling for seven $\alpha$ values ranging from 1.0 (circular) to 3.5 (highly protruding petals). From these shapes, the perimeter $P$ of the solid/fluid domain contact is computed.

The unstrained reference solid domain was modeled as the planar elastic plate, and the elastic energy of solid domains was defined as sum of the strain energy and the bending energy as follows

$$E_{\text{solid}} = \frac{Y}{2(1+\nu)} \int_{\text{solid}} dA \left[ (\text{Tr}\varepsilon)^2 + \frac{\nu}{1-\nu} \text{Tr}\varepsilon^2 \right] + \frac{B}{2} \int_{\text{solid}} dA \, (2H)^2 \qquad (M2)$$

where $Y$ is the 2D Young's modulus, $\nu$ is Poisson's ratio, $B$ is the bending modulus, $\varepsilon$ is the 2D strain, and $H$ is the mean curvature. Whereas the fluid membrane elastic energy was defined by the bending energy only as follows

$$E_{\text{fluid}} = \frac{B}{2} \int_{\text{fluid}} dA \, (2H)^2 \tag{M3}$$

imposing lack of the in-plane shear modulus. The same bending modulus for solid and fluid was assumed for simplicity. We set Poisson's ratio $\nu = 0.4$, and the dimensionless thickness $t/R = \sqrt{B/Y}/R \sim 10^{-4}$ implying relatively much stiffer in-plane strain deformation over out-of-plane bending deformation resulting in nearly vanishing Gaussian curvature over the solid domains, where $4\pi R^2 \equiv A_{\text{solid}} + A_{\text{fluid}}$. We performed the energy minimization simulation in the computational software *Surface Evolver* to minimize the elastic energy for given sets of $\alpha, \phi_s, \bar{v}$ [See Figure 3, example shapes and simulation details in Supporting Information]. The simulation results confirm nearly vanishing Gaussian curvature in the relevant inflation regime, and therefore the strain energy becomes negligible. Since the surface tangent is continuous everywhere including the fluid-solid boundary, the total Gaussian curvature is fixed by topology, that is, $\int_{\text{fluid}} dA \, K_G = 4\pi$, and the equilibrium configurations were found by redistributing $4\pi$ Gaussian curvature over the fluid membranes connecting the developable solid patches minimizing the bending energy. For this reason, the contribution from the Gaussian curvature contribution to the Helfrich bending energy is constant for all configurations and does not contribute to morphology selection for a given solid fraction.

Lastly, to investigate the thermodynamic stability of the solid domain shapes, we considered the line energy

$$E_{\text{boundary}} = \sigma P \tag{M4}$$

where $\sigma$ is the line energy, and $P$ is the boundary perimeter computed from equation (M1). Note that the line energy is dependent on the vesicle size whereas the total bending energy is scale invariant. We combined the line energy with the elastic energy for the values $B = 25 \, k_B T$ and $\sigma = 1 \, k_B T \text{nm}^{-1}$ for vesicles of diameter $D = 30 \, \mu\text{m}$ and compared the total energy among the seven $\alpha$ values in Figure 4.

**RESULTS AND DISCUSSION**

**Experimental Observations**

*Implications of the Phase Diagram for Crystal Mass Evolution*

For the well-studied DPPC-DOPC vesicle membrane system, the established phase diagram[41-43] recapitulated in Figure 1B from our prior work[16] provides a map to strategically navigate crystallization processes. (Figure 1B highlights phase boundaries and other features. Data points themselves are included in the Supporting Information.) An $L_\alpha$ fluid membrane phase containing both lipids coexists with a nearly pure DPPC solid $P_{b'}$ crystalline phase,[41, 44] and constant temperature tie lines, such as the examples shown, bridge the large two-phase region. The ends of a tie line describe the equilibrium compositions of the coexisting $L_\alpha$ fluid and solid membrane phases at the temperature of interest. The proportions of fluid and solid at a given temperature, conforming to a mass balance (termed in some communities as the "inverse lever-arm rule"), depend on the overall membrane composition, described by points within the two phase region.

This study compares crystallization at nominal overall DPPC content of 30, 50, and 70% in the membranes of giant unilamellar vesicles, indicated by three trajectories (vertical lines) in Figure 1B. Cooling at 0.3°C/min produces only one crystal in many vesicles except at 70% DPPC, where the higher density of stable nuclei makes it nearly impossible to find single crystals in larger vesicles, consistent with the known impact of concentration on nucleation density.[39, 40] The impact of cooling rate on solid crystal shape was previously published for 30% DPPC[38] and cooling rate is not examined here explicitly. On all trajectories, vesicles start as a one-phase mixed $L_\alpha$ fluid membrane and, at the highest temperature (we start at 55°C), the vesicles are flaccid.[38] During cooling, thermal contractions of the one-phase fluid membrane produce a spherical vesicle shape before the two phase region is breached.[38]

Importantly, we employ the term "preconditioning" to refer to the evolution of the vesicle tension in the one phase region, indicated for each trajectory in Figure 1B. Then inside the two phase region, crystalline domains nucleate and grow with further cooling, on each of the three trajectories in Figure 1B. At the cooling rates in this study, the fluid membrane composition at any given moment seems, within experimental resolution of visible crystals, to correspond to the equilibrium $L_\alpha$ composition where the tie line intersects the liquidus boundary. If the cooling is halted within the two phase region, the crystal growth also stops, but resumes with further cooling.[39] This observation argues against diffusion-limited crystal growth and against differences between the actual membrane concentration and its equilibrium value during crystal growth. In other words, at each instant in cooling, the composition of the fluid membrane phase approaches that on the tie line.

Figure 1C, which derives from the phase diagram via a mass balance plus molar area data for the fluid and solid phases (with an example calculation detailed in the Supporting Information), shows how the solid area fraction (Top) and relative zero membrane tension equilibrium area (bottom) evolve with temperature along cooling trajectories for membranes of different compositions. The crystallization starts at different temperatures for the different membrane compositions due to the shape of the liquidus boundary and the impact of composition on the melting temperature, seen in the phase diagram itself. At higher DPPC membrane content, there is not only a greater solid area fraction in the membrane at room temperature, but the development of the solid area fraction is nonlinear due to curved shape of the liquidus line. One consequence, evident in Figure 1C, is that for small quench depths into the two phase region, one finds a greater proportion of solid at 70% DPPC compared with the same small quench depth at 50% or 30% DPPC. Indeed a 2°C quench produces, in vesicle membranes containing 70% DPPC, a solid area fraction, $\phi_s$ of ~30%. When the membrane contains 30% DPPC, $\phi_s$ reaches only ~14% at full cooling to room temperature. The bottom part of Figure 1C highlights the effect of composition and temperature on the estimated zero tension membrane area relative to that at 40°C, the approximate

temperature at which the vesicle becomes spherical as opposed to floppy during cooling. The detailed calculations are elaborated in the modeling section. Important to note is the gradual thermal contraction (preconditioning) in contrast to the greater membrane contraction from the phase change itself, once crystallization proceeds.

With an understanding of the evolution of crystal mass for different overall compositions or regions within the phase diagram, it becomes possible to understand the initial and evolving crystal shapes, ultimately in terms of membrane tension. We consider two separate mechanisms for increases in tension, thermal contractions and solidification, (in addition to water permeation-driven tension relaxation). We apply the two mechanisms to cooling trajectories in different compositional regions of the phase diagram, i.e. for high and low concentrations of the crystallizing species, here DPPC. We further break down the experimental observations in terms of 1) differences in initial crystal shape for small quench depths inside the two phase region and 2) the evolution of crystal growth at different composition in the two phase region. Separate consideration of small versus growing crystals helps to reveal the importance of thermal contractions compared with the areal change of crystallization itself.

*Initially Observed Crystal Shape (for Small Quench Depths) and the Impact of Thermal Pre-Tensioning*

The shapes of the initially visible crystals would be expected to be controlled, at least in part, by the state of the membrane (its tension, for instance), at the moment of entry into the two phase region. This is controlled at least in part by the preconditioning part of the trajectory in Figure 1B. Therefore, for 3 different membrane compositions, and following trajectories ending in small quench depths into the two phase region, Figure 1D summarizes the appearances of DPPC crystals as they first become large enough to image as dark shapes within the bright $L_\alpha$ fluid vesicle membrane. Close up images are included in the Supporting Information. The $L_\alpha$ fluid membrane is made fluorescent by trace amounts (< 0.03 mol%) of 1,2- dioleoyl-sn-glycero-3-phosphoethanolamine-N-(lissamine rhodamine B sulfonyl) (ammonium salt) (Rh-DOPE), which is excluded from the DPPC crystals. At the small quench depths indicated in Figure 1C, the individual crystals contribute less the 1-2% of the membrane area (and in the case of multiple crystals the tolal solid contributes less than 4 area %). The crystal morphologies are described quantitatively using a parameter $\alpha$ = outer diameter / inner diameter of the crystals shown: $\alpha$ = 1 for a circular crystal, 1.15 for a hexagon, 1.15-1.8 for a 6-armed starfish, and more than 1.8 for petaled flowers. Along with the plotted summary of crystal shapes, Figure 1D includes micrographs of three typical vesicles at each membrane composition.

Figure 1D suggests a correlation between overall membrane composition, vesicle size, and early crystal morphology. Two regimes are found for the dependence of initial crystal morphology on vesicle size. On small vesicles (~10 μm in diameter), small DPPC crystals always appear as compact with hexagonal faceting sometimes discernible, always independent of composition. On larger vesicles, however, initial crystal morphologies composition-dependent: Crystals tend toward a compact form in membranes with high DPPC content but exhibit flower petal-like protrusions of 6-fold symmetry on vesicles having less DPPC (30%). Viewed differently, Figure 1D demonstrates the key observation that at low DPPC concentrations (but not so low as to produce diffusion-limited crystallization), vesicle size has a profound impact on the morphology of crystals when they first become visible. This trend is much less pronounced at high DPPC content where small crystals tend to be compact independent of vesicle size below about 50 μm diameter. Thus we find a crossover in the shapes of initial crystals suggesting that overall membrane composition impacts preconditioning as the two phase region is approached: For modest DPPC (30%) content, membrane preconditioning imparts a vesicle size dependence to the initial crystal shape. At higher DPPC content (70%) the state of the membrane when the two phase region is breached produces only compact domains.

Notably, this study focuses, as much as possible, on the case where single crystals grow within each vesicle, avoiding the impact of interdomain interactions on the crystal shape. Such interactions can become pronounced for crystals large enough to deform the vesicle membrane. Then the perturbed membrane shape can influence the morphology of growing neighboring crystals. In the case of membranes containing 70% DPPC, we were unable to find vesicles greater than 25 μm that contained only a single crystal. In order to probe the full range of vesicle sizes in the study of small crystal shape Figure 1D includes vesicles with multiple small crystals in larger vesicles containing 70% DPPC. For instance, one 50 μm vesicle containing 70% DPPC, 8 very small crystals of similar size occupied total solid area fraction near 2% with each crystal contributing less than 0.04% to the solid area fraction. We believe these data are reliable because of the smallness of the crystals does not perturb membrane curvature and the total solid area fraction is similar to that in other vesicles. When vesicles contained multiple crystals, they were of similar size and morphology on a given vesicle. We believe that with multiple small domains in a vesicle, the morphology of individual domains is not impacted.

*Observed Morphologies of Growing Crystals*

The final morphologies of crystals depend, potentially, on their initial morphology and on how conditions during crystal growth shift the preference for morphology, for instance compact forms versus those with extended features. Any shift of morphology after initial crystal formation would happen on the "crystallization" part of each trajectory, the black lines in the two phase region of Figure 1B. The observed evolution of crystal morphology, from the initially visible shapes to their final forms at room temperature or until lysis, is summarized for different overall membrane compositions, in the upper sections of Figures 2A, B, and C. (The lower section of each Figures 2A, B, and C includes calculated tensions, described later.) The observed crystal morphologies are indicated by different colors for different alpha values, the ratio of the out-diameter to the in-diameter. The impact of vesicle size is addressed on the x-axes while on the y-axis increasing solid area fraction corresponds to progression through the two phase region. Using solid fraction rather than temperature better enables a comparison across different membrane compositions. Some data, (hollow points) represent sequential images of a growing domain on a single vesicle as it cooled. In other cases, domains on many vesicles were imaged during cooling (solid points). The process histories were identical for both types of data, but more data could be acquired by imaging different vesicles. When viewing plots like Figures 2A-C, tracing an arbitrary vertical upward path keeps vesicle size constant but grows the solid fraction and therefore provides perspective on domain growth on a vesicle of a chosen size.

Comparing parts A, B, and C of Figure 2 reveals the impact of the overall concentration of the crystallizing component (DPPC) on evolving crystal morphology during crystal growth. When the crystallizing component is the minor membrane component, for instance in vesicles containing 30% DPPC, the upper part of Figure 2A reveals that the initial morphology, which depends on vesicle size, is retained as crystals grow upon cooling. That is, alpha is constant as the crystal area fraction, $\phi_s$, increases, ie moving vertically upward, at constant vesicle size, the path one would follow to see crystal growth. In the upper part of Figure 2A, as one follows a vertical trajectory to grow a crystal for a vesicle of a given size, the trace does not cross from one crystal morphology (color) to another. By contrast at elevated DPPC content in Figures 2B or 2C, the morphology of the crystal evolves with its growth. This is evident in the directions of the boundaries between regions of different colors. Thus, a vertical trajectory crosses from one color to another as it progresses upwards, for instance from red ($\alpha \leq 1.15$, compact) to yellow ($1.15 < \alpha \leq 1.8$, 6-armed star) to green ($\alpha > 1.8$, petaled flower ). Thus at elevated DPPC content, crystals that were compact when they were small and first imaged, grow into 6-armed stars and then into flower shapes.

We note that in Figure 2 color boundaries differentiating classes of crystal shapes are sketched with caution due to some overlap between colored points for different vesicles. We believe this blurring the boundaries results from unavoidable vesicle-to-vesicle tension variations that occur naturally due to a few percent variation in volume / area ratios as vesicles are harvested off the electroformer. Nonetheless, the striking composition-dependent boundaries demarking crossovers in crystal morphology are an important feature of the crystallization process. The directions of boundaries in the upper experimental sections of Figures 2A, B, and C, horizontal or vertical, reinforces the impact of composition on the vesicle size dependence of crystal morphology.

Also important to note, the morphological changes during crystal growth appear to be accomplished only by addition of solid to the existing crystal and not to a reshaping (dissolving and subsequent recrystallization) of existing crystals. For instance when flower petals sprout, there is no reduction in size in the solid region that was original compact crystal. Instead there is selective addition of new solid in the forms of petals, and the original material of the compact crystal become the center of a flower. Additionally, for vesicles containing 30% DPPC, when a small flower-shaped crystal grows into a larger one, both the center and petals of the flower become larger. We do not observe a reduction in any part of original solid that would indicate dissolution or rearrangement of crystalized material. It makes sense, then that we observe more pronounced flower petals with lower DPPC content, where small flowers grow into larger flowers. The petals are shorter on vesicles containing 70% DPPC, shown in the scheme of Figure 2D. We believe this is necessarily the case because when protruding shapes grow from compact shapes (as in vesicles containing 70% DPPC), it is not possible to grow a long petal without growing a short petal first. Thus, a larger amount of solid must add to the already grown compact crystal to achieve a flower and sometimes this is confounded by the size limit of the vesicle itself.

Another potentially complicating factor would need to be addressed when solid crystals grow to appreciable fractions of the overall vesicle diameter, towards its equator, where adjacent petals may laterally interact. The shapes of crystals growing over large sections of a spherical surface or the growth of interacting domains involves considerations beyond the current scope.

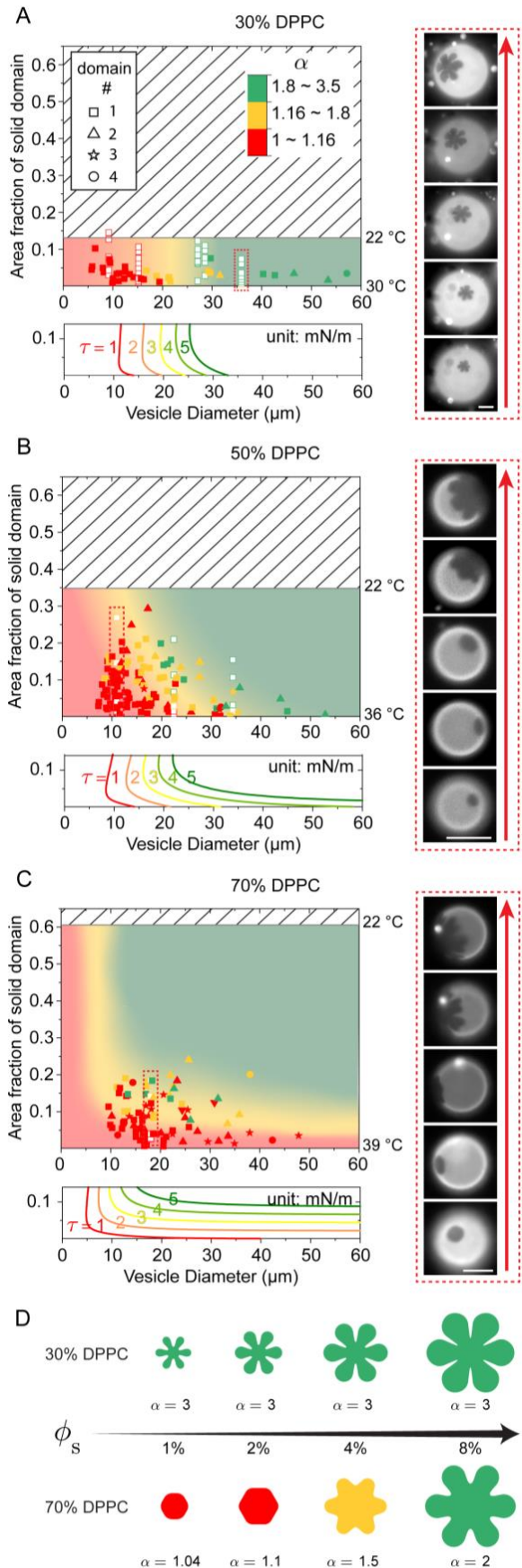

**Figure 2.** Top graphs in A-C for vesicles containing (A) 30%, (B) 50%, and (C) and 70% DPPC: Experimentally-observed evolving morphologies of growing crystals, with different crystal shapes represented by color, as described in the legend of part A: compact (red), star shaped (yellow) and flower-shaped (green) This formatting of data constitutes a processing state space, with upward vertical trajectories tracing the expected evolution of crystal shape, as a crystal grows. The solid area fraction on the y-axis represents the extent of crystal growth for vesicles of different sizes on x-axis. Note temperature axis on right limits solid area that can be achieved at a given membrane composition because cooling ends at room temperature. In (C) for 70% DPPC, experiments terminated upon vesicle rupture. Solid data are for individual vesicles while vertical stacks of hollow points show the evolution of a single crystal. Images on the right of each graph detail crystal growth on example single vesicles. In (A) with 30% DPPC, the original shape was preserved as crystals grew while in (B) and (C) for 50 and 70% DPPC, the crystal evolves protrusions as it grows. Note the vesicle imaged in (C) ruptured after the solid grew to ~20% solid area fraction, as was typical. Most data (squares) are for single crystals on a vesicle; however, where only crystals domains could be found on individual vesicles, the legend indicates the crystal number. In (A), (B), and (C), the lower plots show the curves of constant tension, calculated from the model and recapitulated from Figure 5, with the values of tension indicated and going from red to yellow to green for higher tensions. Note the calculated curves of constant tension, in the lower part of each figure, run in similar directions to the contours in experimental data for evolving crystal shape, in the top part of each figure. (D) Summary of morphological evolution leading to flower morphologies in large vesicles (exceeding 30 μm) containing 30% versus 70% DPPC.

*Summary of Experimental Observations.* We observe important crossover behavior in crystal morphology, as overall membrane composition is tuned from 30% DPPC to 70% DPPC, in terms of the initially visible crystals at shallow quench depths (influenced by preconditioning in the one-phase region) and also in terms of the evolution of crystal morphology (influenced by conditions in the two phase region). At 30% DPPC, the morphologies of initial crystals are size-dependent with compact crystals on small vesicle and more elaborate flowers on larger vesicles. At 70% DPPC, initially small compact domains are preferred on vesicles of all sizes studied. As crystals grow upon further cooling in the two phase region, those membranes comprising 30% DPPC retain crystal morphology upon growth. In the part of the phase diagram where DPPC content is greater, compact crystals evolve into flower forms, a shape shifting behavior that is increasingly pronounced on larger vesicles.

**Modeling**

*How Shear Rigidity Drives Shape Selection Energetics for Small "Primordial" Solid Domains*

For vesicles containing 30% DPPC and solid area fractions near 14%, pairing of experiments with Surface Evolver modeling[16] established that the prohibitive elastic penalty for solid crystals to assume non-zero Gaussian curvature configurations favors flower formation because protrusions (flower petals) can bend cylindrically without imposing large bending cost on the surrounding fluid phase. Modeling further demonstrated that the preferred (observed) crystal morphologies depend on the competition between bending and line energy. For instance while flowers and other morphologies with protrusions avoid Gaussian curvature without large bending cost to the surrounding fluid, their longer perimeters increase the cost of line energy of fluid-solid domain contact. The observed interplay between composition and vesicle size in Figures 1 and 2 suggests further complexity in the mechanism that selects for crystal morphology, especially in the limit of crystal of small area fractions.

To understand how the size of the domain itself influences solid-domain morphology selection, we conducted Surface Evolver modeling, detailed in our previous work,[16, 45] and the current methods section and Supporting Information, for 6-fold symmetrical domains of variable non-convex perimeters and solid area fractions $\phi_s$ of 2, 4 and 8%. Figure 3 shows the elastic energy from Surface Evolver, (including bending energy of both fluid and solid domains and 2D strain elastic energy of the solid) as a function of reduced volume, $\bar{v} \equiv \sqrt{36\pi}\, V/A^{3/2}$ (with $V$ and $A$, the vesicle volume and area) on vesicles where the solid takes on different shapes ( $1.0 < \alpha < 3.5$ ) and area fractions. $\bar{v}$, is a measure of vesicle inflation, approaching unity for spheres. Elastic energy grows large in this limit because having an entirely spherical vesicle would require imposing non-zero Gaussian curvature on the solid domain and generate large and non-zero 2D solid strains. The divergences of the elastic energy shown in Figures 3A-C in fact derive from the growing cost of concentrating bending energy into the fluid phase as a consequence of the expelling Gaussian curvature (and 2D elastic strain) from the solid to the surrounding fluid.[46, 47] These calculations reveal that, for all of the solid area fractions considered, increasing protruding contours of solid domains shapes reduces bending costs relative to those of more compact shapes. Morphologies with protrusions are therefore energetically preferred from this perspective for all values of $\bar{v}$, increasingly so at larger inflation.

Comparing Figures 3A-C, it is clear that bending energy costs are smaller in vesicles whose solid domains comprise a smaller fraction of the total area and further, that for smaller solid domains compared with larger ones, the points at which elastic energy is minimal and where it diverges shift to higher inflations (i.e. closer to $\bar{v} \to 1$). Hence, for small, "primordial" domains compared to the large solid domains that are ultimately achieved, the elastic energy drive that favors flower formation sets in at higher inflation (i.e. $\bar{v}$ value). For small domains, a small degree of under-inflation effectively eliminates the bending-energetic incentive to assume a non-convex flower shape in comparison to the generic tendencies of line energy to favor compact, convex shapes. However as domains grow larger, inflation becomes increasingly costly: For a given vesicle inflation, i.e. a given $\bar{v}$, growth of a solid domain, which starts small, will eventually grow so big that it could sprout protrusions to lower its bending energy.

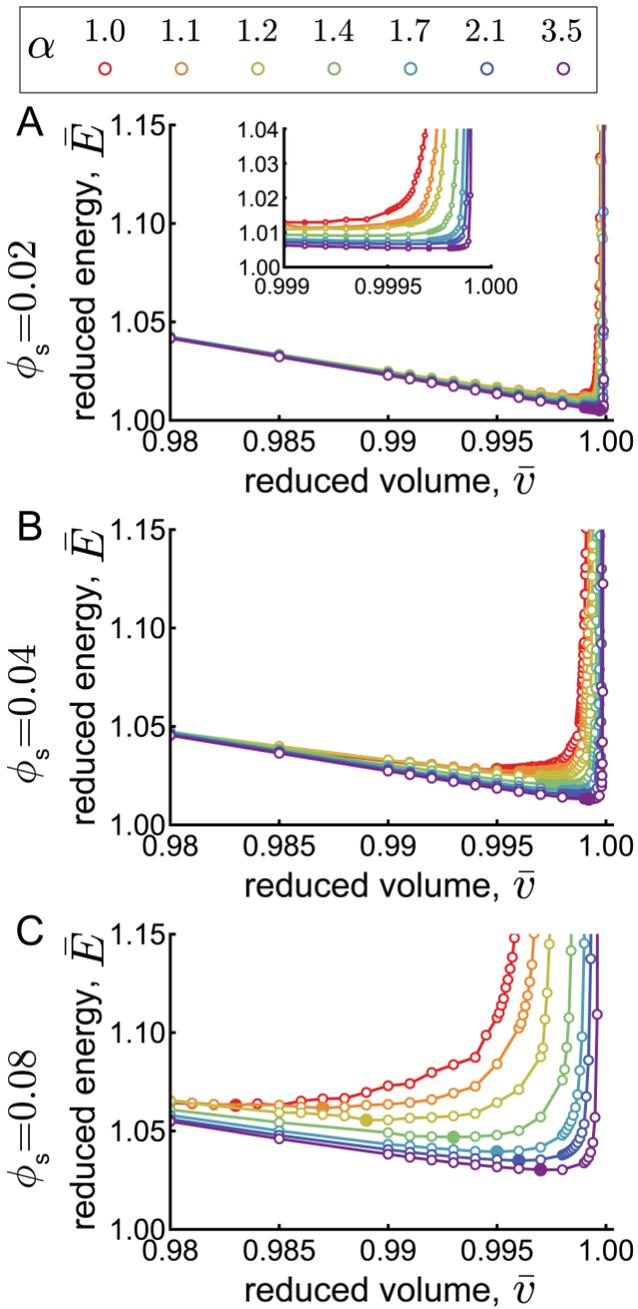

**Figure 3**. Elastic energy, in reduced form scaled on that for a fluid sphere, ($8\pi B$), as a function of reduced volume. Vesicles contain solid area fractions of A) 0.02 B) 0.04 and C) 0.08. The shapes of the single solid domain are indicated by the colors, in the legend at the top.

We model this effect by combining the elastic energy with the product of the domain perimeter and the line energy cost per unit length, $\sigma$, to compute the total energy of the composite vesicle as function of $\bar{v}$ for various shapes and solid domain sizes, shown Figure 4. (For comparison here we consider a fixed vesicle radius for all three solid area fractions; with scaling on vesicle size in the Supporting Information Figure S4.) Note that the line energy cost is independent of inflation, as the perimeter is fixed for a given solid shape and size. However, the overall perimeter increases with the (square root of) solid domain fraction, leading to an increasing contribution from boundary energy (curves in Figure 3) for larger solid domains, reflected as larger vertical shifts for 8 and 4% solid fraction relative to 2%. Additionally, for a given solid domain and vesicle size, the vertical shifting of total energy versus $\bar{v}$ is greater the for more elaborately protruding, and larger perimeter flower domains, countervailing the tendencies of the elastic energy to favor highly petaled domains.

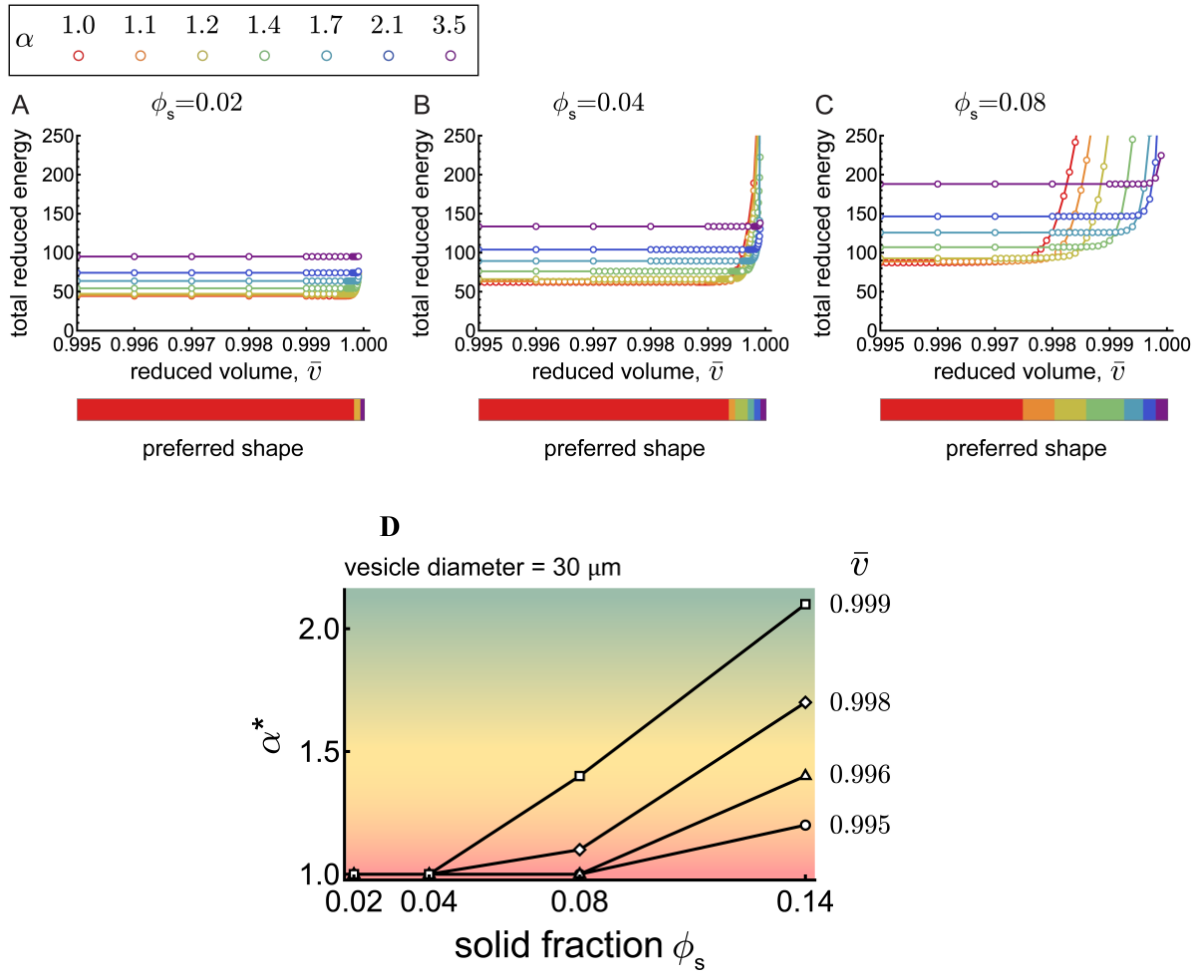

**Figure 4**. For vesicles containing a single solid domain having the area fraction indicated A) 0.02, B) 0.04 and C) 0.08, the plots show the combined bending and line energies (reduced) for domains of different shapes (colors) in vesicles of different reduced volume, (x-axis). The total reduced energy has units of bending stiffness. The color bar below each plot shows preferred domain shape for different reduced volumes corresponding to axis in each figure. (D) Shows the preferred domain shape as crystals increase in size (solid fraction), for fixed reduce vesicle volumes. Points are based on the total reduced energies in parts A-C, and lines between points simply guide the eye. Because simulations are conducted for a finite selection of alpha values, the results for preferred shapes are discretized and limited to these values.

The lowest total energy, combining elastic and boundary energy determines the conditions for energetic morphology selection (i.e. compact vs. petaled solid shapes) on vesicles of a given inflation $\bar{v}$ and for a series of increasing solid fractions, in Figure 4A-C. This representation reveals that with a solid domain comprising 2% of the overall area, nearly all of the state space favors compact domains (the lower panels along Figure 4A-C are mostly red). Thus simulations suggest that as long as a domain is small enough, it should be compact unless the vesicle is nearly at full inflation (i.e. a perfect sphere). As a domain grows larger, the range of reduced volumes where compact domain morphologies are selected (red area) diminishes in favor of shapes with increasing protrusions. Increasingly elaborate flower forms develop as a crystal grows, with $\phi_s = 0.02, 0.04$ and $0.08$, in parts A, B, and C of Figure 4. Figure 4D summarizes the shifting of the preferred crystal morphology with increased crystal size (area fraction) at fixed inflations. Ultimately for sufficiently large crystals, complex shapes are generically preferred unless the vesicle deflates considerably during the process, for instance via rupture or by water permeation.

The transformation anticipated in Figure 4D from compact to flower forms is seen in the experimental upper plots Figure 2 in vesicles of moderate size for DPPC content of 50 and 70%. The observation of small flowers in vesicles of 30% DPPC suggests that the transformation from compact to flower-shaped crystals takes place before crystals are large enough to be imaged, suggesting vesicles at 30% DPPC are more inflated during crystallization, or at a higher tension state than vesicles with greater proportions of DPPC. This impact of membrane composition on inflation and tension, and the interplay of these variables with vesicle size, is addressed next.

*Modeling Tension Evolution*

In order to understand how membrane composition impacts the evolution of crystal morphology, and how the competition between elastic and line energy, which selects for crystal shape, can be exploited through process variables, we consider the evolution of membrane tension. Membrane tension is the Legendre transform of the total bending energy with respect to $\bar{v}$, i.e. the reduced volume derivative of the bending energies (i.e. slopes of curves) of Figure 3.

*Factors Increasing Membrane Tension.* Membrane tension is increased as a result of membrane contractions from both thermal and phase transition processes. During "preconditioning," the cooling that occurs before the system enters the two phase region, thermal expansivity-related membrane contractions[38] drive an increase in membrane tension. The tension starts to increase once a vesicle, originally floppy (zero tensioned) and underinflated at elevated temperatures, contracts thermally to the point of becoming spherical, observed to occur around 40°C. Thermally-driven tension increases can continue with further cooling in both the one and the two phase region.[38] Crystallization in the two phase region causes a further reduction in the equilibrium vesicle area that that further increases membrane tension.

We address these details quantitively by building from a previous model which considered only thermal expansivity-driven membrane contractions to increase tension, countered by tension relaxation via water diffusion out across the membrane.[38] Now, in a more general treatment, the additional impact of the crystallization phase change, which dominates when the membrane contains greater proportions of DPPC and greater ultimate solid area per Figure 1C, is taken into account.

When the temperature of a single-phase fluid or solid membrane is changed, a coefficient of thermal expansion, $\kappa$,[12] defined in equation 1A, describes the change of its equilibrium area, $A_{eq}$:

$$\kappa \equiv \left[\frac{1}{A_{eq}}\frac{\partial(A_{eq})}{\partial T}\right]_{\tau=0} \tag{1A}$$

We employ this definition in the zero tension limit, where literature data for both fluid and solid phases are available, treating $\kappa$ as constant.[48] Such zero tension area changes are evident when a fluid ($L_\alpha$ phase) vesicle is underinflated (floppy) during a temperature change, enabling $A_{eq}$ to be measured, for instance by low suction micropipette aspiration.

When molecules in a fluid membrane transform into a solid membrane phase, the difference in equilibrium (zero tension) molecular area in the fluid and solid phases imparts a contraction: $\Delta A^{f-s}$ per mole. When the membrane contains a single species, crystallization occurs at fixed $T_m$; however, when membrane comprises multiple phospholipids, the liquid composition travels along the liquidus line as the temperature decreases, in Figure 1B. For an incremental reduction in temperature $dT$, the mixed membrane experiences an incremental increase in solid area fraction, $d\phi_s$ and undergoes an incremental contraction from the change in state, along with the thermal contraction of the fluid and the thermal contraction of any solid phase already present. We write these combined effects of temperature on the equilibrium membrane area as:

$$\frac{dA_{eq}}{A_{eq}} = \left[\kappa_F\big(1-\phi_s(T)\big) + \kappa_s\phi_s(T) - \Delta A^{f-s}\frac{d\phi_s(T)}{dT}\right]dT \tag{1B}$$

Here, $\kappa_F$ and $\kappa_s$ are the thermal expansivities of fluid and solid membrane domains and the term in square brackets includes both thermal and phase-change contributions to the reduction of the membrane area. We define

$$\kappa' \equiv \kappa_F\big(1-\phi_s(T)\big) + \kappa_s\phi_s(T) - \Delta A^{f-s}\frac{d\phi_s(T)}{dT}, \tag{1C}$$

which includes the relative amounts of membrane contraction from thermal effects versus the phase change. In the expression 1B and Figure 1C, $\kappa'$ is not quite constant, but since $\kappa_s$ is nearly equal $\kappa_f$,

$$\kappa' \cong \kappa - \Delta A^{f-s} \frac{d\phi_s(T)}{dT}$$ At 30% DPPC $\kappa$ dominates. However at 70% DPPC, the second term dominates. This composition dependence arises because, as shown in the top part of Figure 1C, a large amount of solid is produced for a small temperature change at an overall composition of 70% DPPC, but the opposite is true at 30% DPPC. These differences are a result of the shape of the liquidus line of the phase diagram.

The evolution of the equilibrium (zero tensioned) area $A_{eq}$ with temperature, is summarized in the bottom part of Figure 1C, by integrating equation 1B from 40°C to lower temperatures. Here we employed the solid area fractions from the top of Figure 1C and physical properties estimated from the literature in Table I. An arbitrary reference temperature of 40°C is selected, corresponding to the approximate temperature where floppy, deflated vesicles appear spherical around 40°C.

The bottom of Figure 1C reveals how the calculated equilibrium (zero tension) area, $A_{eq}$ initially decreases via thermal contractions and then more sharply with crystallization. For vesicles containing 70% DPPC, a greater ultimate amount of crystallization leads to more extensive membrane contractions. With 70% DPPC, the initially large amount of crystal formed per cooling increment in the two phase region produces an abrupt contraction on top of the thermal contraction. For vesicles containing 30% DPPC, a more gradual reduction in $A_{eq}$ is calculated as a result of continued thermal contraction, and due to the protracted trajectory in the one phase region the overall area reduction is substantial before crystals nucleate. Thus the membrane contraction associated with the phase change is more important for membranes containing 70% DPPC compared with 30% DPPC because for a given quench depth, more crystallization occurs for the former, a result of the shape of the phase diagram.

Table 1 Estimated physical properties employed in models

| Physical parameter | Estimated value | References |
|---|---|---|
| Thermal area expansivity for both solid and fluid phase $\kappa_S, \kappa_F$ | 0.005 °C$^{-1}$ | 48 |
| Area expansion modulus $K_A$ | 237 $mN\ m^{-1}$ | 49 |
| +Water permeability of bilayer membrane $\mathcal{P}'$ | 42 $\mu m\ s^{-1}$ | 50 |
| Fractional area changes during phase transition from fluid to solid phase $\Delta A^{f-s}$ | 0.17 | 48 |
| Interfacial area per lipid molecule for solid phase (mainly DPPC) $A_S$ | 46.4 Å$^2$ | 51 |
| Interfacial area per lipid molecule for fluid phase $A_F$ | 69.6 Å$^2$ | 51 |
| Bending modulus $B$ | 25 $k_B T$ | 49, 52 |
| Line energy of solid-fluid phase boundary $\sigma$ | 1 $k_B T\ nm^{-1}$ | 53 |

+ See supporting information for relation of $\mathcal{P}'$ to $\mathcal{P}$, defined in equation 4.

This change in equilibrium area in Figure 1C is not always accessible in experiments. For instance, when a vesicle membrane is constrained by the volume of a vesicle's contents, the actual membrane area $A$, will likely exceed $A_{eq}$, leading to finite membrane tension, $\tau$:

$$\tau = K_a \left(\frac{A - A_{eq}}{A_{eq}}\right) = K_a (A' - 1) \tag{2A}$$

Here, $K_a$, is the area expansion modulus and $A'$ is the areal deformation, $A/A_{eq}$.

While the tension vanishes and the vesicle is floppy, $A = A_{eq}$. Once the vesicle is spherical, however, further cooling reduces the equilibrium area according to equation 1B, and the volume constraint causes the tension to grow, still independent of vesicle size.

*Factors Relaxing Tension.* When the vesicle membrane is tensed, the pressure inside the vesicle exceeds the exterior pressure as described by Laplace:

$$\Delta P = \frac{2\tau}{R} \qquad (3)$$

where R is the vesicle radius. Since phospholipid membranes are somewhat water permeable, this pressure difference from the inside to the outside of the vesicle drives water out of the vesicle, and produces a volume reduction:

$$\frac{dV}{dt} = -A \frac{\wp}{\rho} \Delta P \qquad (4)$$

Equation 4 defines the membrane permeability, $\wp$ which may be a combination of crystal and fluid membrane permeabilities: $\wp = \phi_s \wp_s + (1 - \phi_s) \wp_f$ and $\rho$ is the density of water. Note now, in equation 3 and 4, vesicle size has entered the model.

*Overall Tension Evolution*. For vesicles that are spherical, $V$ and $A$ are related such that $dV = \frac{1}{4\sqrt{\pi}} A^{1/2} dA$. Even with a solid domain present, this relationship between volume and area is a good approximation. Then, equations 1B through 4 can be combined to yield:

$$dA' = -\left[k''(A' - 1) + A' \left(\frac{dT}{dt}\right) \kappa'\right] dt \qquad (5)$$

This form explicitly shows the impact of transport (the first term in square brackets) and thermal and crystallization-related contractions. Thus the tension evolution is controlled by two groups, $k''$ and $\left(\frac{dT}{dt}\right) \kappa'$, with $\kappa'$ discussed above. The time constant

$$k'' = 16 \pi \frac{\wp}{\rho A_{eq}} K_a, \qquad (6)$$

combines the membrane's water permeability and area expansion modulus. It also contains, grouped with the permeability, the equilibrium area of the membrane (available for water transport), which changes with cooling by only a few percent but varies considerably for vesicles of different sizes. It should be noted, however, that changes in $A_{eq}$ at this position of the equation with time are small and have a minimal influence on its solution, no more so than minor (a few percent) changes in the permeability or area expansion coefficient, which are treated as constant. The impact of cooling on these physical property data, or the time evolution of $A_{eq}$ of that matter, would constitute a further refinement of the model and so the quantities are grouped, in the current treatment into constant $k''$. Indeed, variation in permeability due to crystallization is negligible for shallow quenches and small domains, and contributes little for all the data (20%) compared with the impact of vesicle size on $k''$. Worth noting for the physical properties in Table I, $k''$ takes on values of 0.0116, 0.0013, and 0.00046 s$^{-1}$ for vesicles of diameters 10, 30 and 50 µm, respectively detailed in Section 5 of the Supporting Information. The effective relaxation time, $k''^{-1}$ varies from 86 s for 10 µm vesicles to 36 minutes for 50 µm vesicles. These time constants make the impact of stress relaxation easy to discern on the laboratory time frame.

Without crystallization, $\phi_s = 0$ and equation 5 can be solved analytically.[38] However, incorporating the shape of the liquidus line of Figure 1B and the mass balance leading to the curves of Figure 1C for $\phi_s(T)$ requires that equation 5 be solved numerically, as was done in this work.

Equation 5 was solved using Mathematica, employing the physical property data of Table I, a polynomial fit to the measured phase diagram boundaries of Figure 1B, and $\phi_s(T)$ shown in Figure 1C. Based on qualitative observations of heated vesicles which appeared floppy at elevated temperatures in the one phase region but which regained a more taut circular appearance near 40°C, vesicles were set to zero-tensioned one-phase spheres at 40°C and, upon further cooling experienced increases in tension and areal strain. Figure 5 summarizes the tension evolution for membranes of different compositions. Each Figure 5 section, A, B, and C, corresponds to a particular vesicle diameter. The necessity of grouping data by vesicle diameter emphasizes its importance. The shapes of the initially visible small crystals at shallow quenches are drawn in schematically for reference.

In Figure 5 for a given vesicle size, membranes of different compositions initially travel on a common tension trajectory during cooling through the 1-phase L$_\alpha$ fluid region, but then branch individually upward towards elevated tensions as the two phase region is breached at composition-dependent temperatures.

This behavior is anticipated in Figure 1C: Starting when vesicles first become spherical near 40°C, cooling trajectories at different compositions undergo different further extents of cooling prior to reaching the liquidus line. Vesicles having 70% DPPC enter the two phase region earlier and at higher temperatures than those whose membranes contain less DPPC. In Figure 5, the underlying tension trajectory from only thermal contractions is continued as a blue-gray line, showing the impact of tension if crystallization did not occur.

In Figure 5, the tension trajectories upon cooling exhibit dramatically different shapes depending on both vesicle size and composition. For instance, with 70% DPPC, vesicles experience a sharp rise in tension upon entry into the two phase region. This occurs as a result of the sudden large amount of crystallization that occurs for in an incremental change in temperature, evident in Figure 1C, a result of the gradually sloped liquidus line in the neighborhood of 70% DPPC, in Figure 1B. By contrast, the rise in tension experienced by vesicles containing 30% DPPC upon entry into the two phase region is more gradual as a result of the steeper liquidus line near 30% DPPC, producing a small around of crystallization for the same cooling increment in Figure 1C. Important to note, however, is that because vesicles containing 30% DPPC undergo a longer cooling step before entering the two phase region, their tension at the time of nucleation is higher than it is for 70% DPPC. Thus it is evident that the amount of crystallization for a given $\Delta T$ (in Figure 1C and $\kappa'$ in equation 1B) translates to different amounts of tension increase in Figure 5. Further, for example, the membrane contractions from the phase change dominate for vesicles containing 70% DPPC, whereas the phase change has a small effect on tension when vesicles containing 30% DPPC, where tension is controlled by thermal contractions.

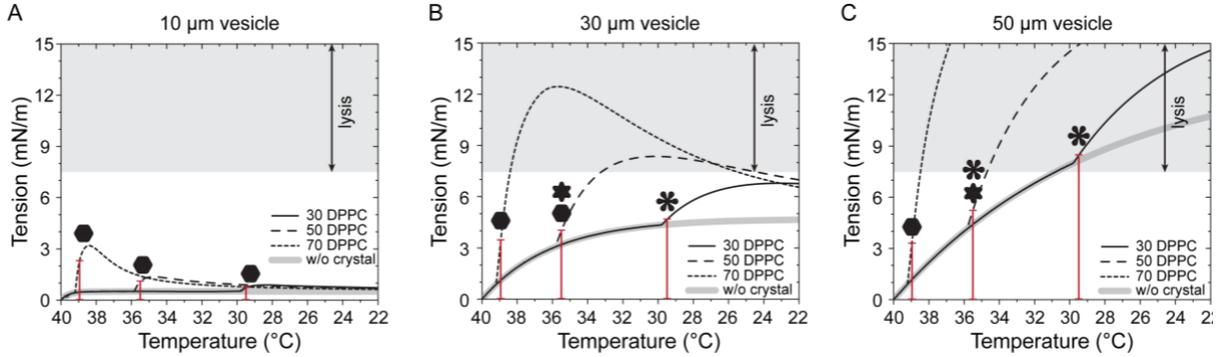

**Figure 5.** Calculated tension histories during cooling, for (A) 10μm, (B) 30 μm, and (C) 50 μm vesicles of different membrane compositions, indicated. Crystal shapes indicate the shapes of the initial experimentally observed crystals (in Figure 1) following a shallow quench to produce a solid area fraction of 1-2%.

*Simple Generalized Expression for Crystallization in Membranes of Any Composition.* While the expression in equation 5 can be solved for any membrane system, employing physical properties and a numerical approximation of the liquidus line for any new system, a simplified expressions for the time, temperature, and magnitude of the tension overshoot would allow a direct extension of the tension model to other system and provide additional physical insight into the transient tension behaviors we report. We detail the development of the simplified expressions in Section 6 of the Supporting information: Specifically the characteristic time, $k'''^{-1}$ (in equation 6 of the main paper) approximates the time from entry into the two phase region to the maximum tension. Then, a linearized treatment anticipates the depth of temperature quench, $\Delta T_{max}$, corresponding to the peak tension:

$$\Delta T_{max} = \frac{dT}{dt} k''^{-1} \qquad (7)$$

Note here, that $k'''^{-1}$ depends on membrane properties and vesicle size, but not explicitly on composition. Therefore $\Delta T_{max}$ exhibits only a weak composition dependence. For a 30 μm vesicle having $k'''^{-1}$ = 12.8 min, a cooling rate of 0.3 K/min gives a $\Delta T_{max}$ between 3 and 4 K, in agreement with Figure 5B and Figure S6B.

We also developed an expression for the height of the overshoot, linearizing the initial rise of the tension inside the two phase region and multiplying by $\Delta T_{max}$:

$$\tau_{max} = \Delta T_{max} \left[\frac{\partial \tau}{\partial T}\right]_{\phi_s \to 0} \tag{8}$$

Working through the math yields the following expression:

$$\tau_{max} = \left(\frac{dT}{dt} k''^{-1}\right) K_a \left(\Delta A^{f-s}\right) \frac{c}{S_{liq}(z_A - 1)} \tag{9}$$

Here the quantities include the cooling rate, the time constant, $k''^{-1}$, from equation 6, the area expansion modulus of the membrane $K_a$, the fractional membrane area change (a contraction) upon crystallization $(\Delta A^{f-s})$, the ratio of the solid to fluid membrane molar areas, $c$, and the slope of the liquidus line on the phase diagram, $S_{liq}$, at the overall concentration of crystallizing species, in the membrane, $z_A$. Worth noting, $c$ was used in the lever arm rule to calculate the solid area fraction in Figure 1C from the phase diagram in Figure 1B, as described in Section 1 of the Supporting Information. Thus for crystallization of a membrane component upon cooling into a two phase region of the membrane, knowledge of a handful of physical properties along with the shape of the phase diagram anticipates the maximum tension.

For our system, as an example detailed in the SI, for a 30 μm vesicle comprised of 70% DPPC equation 9 predicts $\tau_{max} = 34\ mN/m$, which exceeds the solution to equation 5 in Figure 5B, due to the linearized form in equation 8. However, equation 9 predicts $\tau_{max} = 2.9\ mN/m$ above the underlying thermal tension rise for vesicles containing 30% DPPC, which is a reasonable approximation. Notably the maximum tension anticipated by equation 9 predicts reasonably when tenson would become so large as to produce lysis.

Moreover, equation 9 predicts that for vesicles of a given size, but different compositions, $\tau_{max}$ scales inversely as the slope of the liquidus line and $z_A - 1$. That is, where the boundary of the two phase region is flatter, greater tension overshoots will be observed, here at 70% DPPC. At 30% DPPC, the slope of the liquidus line in Figure 1B is greater and as a result the tension overshoot is not as great.

**Pairing Modeling and Observations**

*From Tension Preconditioning to Initial Crystal Shape.*

The morphologies of the experimentally observed crystals are indicated schematically on the individual tension history curves of Figure 5 for small quenches into the two phase region, to provide perspective on the results in Figure 1D.

For small vesicles, where water permeation is sufficiently fast to maintain a low tension throughout the cooling and crystallization process, the low tension correlates with the observation of initially compact domains at the time of their first observation (Figure 1D. This is the case in membranes containing both small or large amounts of the crystallizing component, DPPC.

For large vesicles, the tension reduction by water diffusion is slower than it was for small vesicles such that accumulated tension can overshoot and does not immediately relax. This produces an impact of membrane composition on tension history. For instance, in large vesicles containing 30% DPPC, the long cooling trajectory and associated membrane contractions in the one phase region precondition an elevated membrane tension before the two phase region is reached. By contrast in large vesicles with 70% DPPC, a shorter cooling trajectory in the one phase region does not produce as high a preconditioned tension upon entry into the two phase region. These tension differences at the start of crystallization are evident in the traces of Figure 5 and are found to correlate with the observed initial crystal morphologies, seen in Figure 1D and shown schematically in Figure 5. It is the shape of the liquidus line, and the lower crystallization temperature with lower amounts of DPPC, that determine the extent of preconditioning and the initial crystal morphology, when crystallization occurs before tension relaxes via water permeation.

Striking differences in the tension trajectories also stem from the vesicle size-dependent tension relaxation rates, k", when water permeates out from vesicles, opposing the effects thermal and crystallization-driven membrane contractions. The time constant for water permeation-driven relaxation, 1/k", scales as the vesicle size squared, such that larger vesicles relax much more slowly (36 min for a 50 μm vesicle) than do small vesicles (86 s for a 10 μm vesicle), a result of their relatively small surface to volume ratios. Thus composition-dependent accumulated tension is slower to relax to zero on large vesicles. High tensions, resulting from thermal contractions or the phase change itself, favor crystals with protruding rather than compact shapes. Especially on large vesicles where membrane tension can grow faster than it relaxes, membrane tension can become extremely high in the absence of other factors. Phospholipid membranes, however, are limited in their ability to maintain high stress, lysing at tensions in near 7-10 mN/m[50] and resealing in a cyclical fashion[54] that places a practical upper bound, highlighted in Figure 5, on tension and introduces complexity in tension history. It is not clear for each burst, how much tension is lost before resealing occurs but many such cycles can occur and potentially perturb thee shapes of growing crystals.

*A Processing State Space for Evolving Crystal Shape from Predicted Lines of Constant Tension*

The calculated tensions, represented as tension histories in Figure 5 as a function of temperature during cooling for vesicles of different sizes and compositions, are replotted as a function of vesicle size for different vesicle compositions in the bottom parts of Figure 2. Comparing the upper parts of each of Figure 2A, B, and C to the lower parts of each enables a comparison of experimentally observed domain shape and calculated tension, respectively. In this representation, the calculated lines of constant membrane tension (lower part of each figure) are found to run parallel to boundaries between experimentally measured crystal shapes (upper part of each figure), establishing the connection between crystal morphology and membrane tension (or inflation), as suggested in the Surface Evolver Calculations of Figure 3. Particularly striking is the influence of membrane composition which alters the curvature and direction of the lines of constant membrane tension and boundaries between crystal morphologies. Specifically in Figure 2A for vesicles containing 30% DPPC, lines of constant membrane tension are mostly vertical as are demarcations in the data between crystal morphologies. With a progression in Figures 2B and 2C to 50 and then 70% DPPC, the constant membrane tension lines and the boundaries between crystal morphologies develop substantial horizontal character, especially for larger vesicles. Thus starting with a vesicle of a select size, on the x-axis of Figure 2, one moves vertically upward through the processing space as a crystal grows. With vesicles containing 30% DPPC, the tension is nearly constant starting from the time of nucleation and throughout crystal growth (because it had increased during cooling in the one-phase region), approaching $\phi_s = 0.14$ at room temperature. Conversely with higher DPPC content and especially for large vesicles, growing a crystal from zero to finite solid fraction requires a vertical upward trajectory to cross lines of constant tension. Thus the membrane experiences a tension increase with increased solid content. This tension increase and shape evolution is evident in Figure 2C for membranes containing 70% DPPC , which $\phi_s \sim 0.50$ at room temperature. In reality, membrane tension rises so high that vesicles ruptured and experiments terminated near $\phi_s \sim 0.25$.

*From Membrane Tension to Crystal Morphology.* Experiments and theory uncover the competing mechanisms that underlie the impact of composition on the shape evolution of 2D crystals formed at shallow quenches and upon cooling to room temperature: elastic versus line energy and composition-dependent stress growth (including thermal versus solidification terms) versus water permeation - associated stress relaxation. Though complex, experimental observations and membrane tension predictions produce a processing state space, in Figure 2, which anticipates the class of crystal shape and its evolution.

The finite element calculations in Figure 3 provide insight into the relationship between membrane tension and crystal morphology: The slope of a given curve, at a point corresponding to a reduced volume of interest, is proportional to tension since tension is the Legendre transform of bending energy. Thus from Figure 3 we can qualitatively discern that, for a given reduced volume (vesicle inflation), growth of a single compact crystal produces a greater tension in the membrane than does a flower shaped crystal. That is, sprouting of petals from a compact domain reduces membrane tension as bending costs are reduced. It follows then, that if tension is forced to be high, for instance from thermal contractions and slow water release, that flowers will be the preferred morphology rather than compact crystals. These tendencies explain the correlation between the observed experimental shapes and the tensions predicted by the engineering model, for instance the parallels in Figure 2. Indeed we have previously confirmed elevated tensions on large vesicles 20 minutes after cooling[16] and here observed rupture of large vesicles, especially at 70% DPPC, preventing quantification of crystal shape at room temperature in the latter, consistent with the predictions in Figure 5.

*Small Crystal Limit*. While we do not argue that nucleus morphology is tension-dependent, Surface Evolver modeling, down to $\phi_s = 0.02$, suggests no limit in $\phi_s$ for the preference for flower-shaped domains when membrane tension is elevated before crystallization, i.e. prior to crossing the liquidus line during a quench. However, the trends shown in Figure 3 and the lack of a mechanism to diminish the sharp upturn in bending energy with inflation at small $\phi_s$, suggests that when vesicles are at tense or approach a spherical shape, flower shapes will be preferred and therefore might form well before crystals can be imaged. This possibility might seem counterintuitive: Giant unilamellar vesicles, spheres on the order of tens of microns in diameter, are considered to be locally flat. However, small but non-zero amounts of Gaussian curvature are incompatible within solid crystals, push Gaussian curvature to the $L_\alpha$ fluid regions of the membrane and profoundly impacting morphology. Giant unilamellar vesicles, as locally flat as they seem by some measures, still possess ample curvature alter crystal growth.

*Towards Generality: Combinations of Variables*

Beyond the linearized generalizations in equations 7-9, the Surface Evolver calculations and membrane tension modeling both reveal combinations of variables that define behavior. The competition between elastic and line energy, critical to crystal shape selection in Figure 4, is borne out in the reduced vesicle size in the Supporting Information, $R/(B\sigma^{-1})$, where $R$ is vesicle radius, $B$ is bending stiffness, and $\sigma$ is line energy. While a vesicle diameter of 30 μm was selected for Figure 4, the Supporting Information shows the insensitivity of the selected domain shape this choice. In fact the results of Figure 4 hold for a wide range of parameter values, as long as the line energy is strong enough to drive compact domain formation, which it usually is for phospholipids and many other systems. Thus the dependence of crystal shape on vesicle size originates, in experiments, through other means.

The membrane tension model reveals two other important parameter groups, $\kappa' \frac{dT}{dt}$ and $k''$. $\kappa' \frac{dT}{dt}$ represents a fractional rate of membrane shrinkage with time. Here $\kappa'$ combines thermal and phase-change driven mechanisms, but more generally, $\kappa' \frac{dT}{dt}$, as a rate of shrinkage is a general processing property of a 2D material, especially one that is annealed thermally. Different physics for other systems could be formulated and substituted. This study employed a single fixed cooling rate of 0.3 °C/min, a convenient laboratory rate on which to conduct experiments, and it was demonstrated that here a large variety of crystal shapes were accessible. A previous work[38] examined the impact of cooling rate for the specific composition of 30%DPPC / 70% DOPC, establishing that slower cooling, which is a difficult experiment, reduced tension. The model presented here includes the ability to change the cooling rate, but the quantitative experimental study at elevated DPPC concentrations is left to future work. Worth noting, $\kappa' \frac{dT}{dt}$ does not introduce curvature or vesicle size.

The final quantity, the time constant $k''$, incorporates several physical properties in addition to the vesicle area, which is where vesicle size enters the physics. With the $R^2$ dependence of the time constant, and variations in vesicle size by factor of 5 or more in experiments, vesicle size dominates the dynamics of stress relaxation relative to that imposed through the cooling rate $\kappa' \frac{dT}{dt}$. While the membrane permeability affects $k''$ to first order, even a factor of two change due to shifting of fluid phase composition during crystallization, or through the appearance of a water-impermeable crystal domain that is 10-20% of the surface area, will have a small impact on $k''$. Other details like the impact of temperature on physical properties are also expected to be less important than vesicle size. For instance, $k''$ contains the product of permeability and modulus $K_a$, such that an increase in temperature will increase permeability but decrease modulus, reducing the overall effect of temperature on $k''$. Most importantly towards the generalization of a tension model to broad systems for processing of 2D materials, it is important to understand the physical origin of stress development and relaxation in other systems, to develop analogous time constants. By considering these quantities and linearizations for the estimated tension maximum, we provide means to apply our treatment to crystallization in other membranae systems.

**CONCLUSIONS**

We demonstrate how development of small 2D phospholipid crystals and their growth upon cooling within the otherwise fluid membranes of giant unilamellar vesicles is strongly influenced by overall membrane composition, through the position of the composition on the phase diagram relative to the fluid-solid boundary and the slope of that boundary, which dictates the relative contributions of thermal- and solidification-driven membrane contractions. This altered membrane tension tunes the trade-off between elastic and line energy to select for initial and growing crystal shape.

When the membrane area fraction of crystalline solid produced per increment of cooling is relatively large, which occurs when the concentration of the crystallizing species is relatively high, then membrane stress is dominated by the growth of areally-dense crystallized solid relatively to the less dense membrane fluid. The membrane contraction associated with the phase change drives an increase in membrane tension, especially for large vesicles whose surface area to volume ratio is insufficient to allow enough water permeation for ample stress relaxation. Then initially compact phospholipid crystals develop protrusions as they grow, in order to accommodate the overall all Gaussian curvature of the inflated highly tense vesicle. Such processes are less pronounced on smaller vesicles because water permeation from the vesicle prevents accumulation of high membrane tension and reduces vesicle inflation relative to that of a sphere (reduced volume).

In the opposite limit where the crystallizing membrane component is more dilute (but not so dilute as to produce diffusion-limited crystallization), the membrane area fraction increase of the crystal per increment of cooling is smaller and membrane stress results predominantly from thermal contractions which occur even before crystals nucleate. Then by the time small crystals become visible in an optical microscope, they already display a morphology dependence on vesicle size, with shapes with protrusions (flowers) reflecting the higher sustained stress on larger vesicles where accumulated water permeation is slower. By comparison on small vesicles with the crystallizing component being the minor membrane species, thermal stresses are relieved on processing timescales by water permeation and initially visible domains are compact and remain compact during their growth.

In this way the shape of the phase diagram and the overall membrane composition influence either the initial crystal morphologies when the crystallizing component is the minor species, or the morphology evolution during crystal growth, (but not the initial morphology), when the crystallizing component is the major membrane species. The impact of vesicle size results from the $R^2$ time dependence of water transport-related membrane tension relaxation. The total membrane tension connects to the relative bending and line energy costs to form crystals of different morphologies on vesicle with different degrees of deflation.

The work demonstrates the complex interplay of thermodynamic features, materials behaviors, and kinetic timescales on control of 2D crystal morphology. Understanding the connection between inflation, membrane tension, and preferred crystal shape, and controlling membrane tension through an understanding of contributing factors evident in the shape of the phase diagram enables the design of processes that can target crystals of desired shapes and sizes, using lamellar membrane templates. While the results here employed the DOPC-DPPC system, modeling suggests the generalization of this behavior to crystallization other membrane systems, inputting separately determined physical membrane properties that do not place restrictions on the system at the molecular level. Both the "engineering model" that anticipates tension does so reasonably with input of membrane permeability and modulus estimates; while the Surface Evolver treatment plus line energy model requires estimates of edge energy and membrane mechanics.

*Conflicts of interest*
The authors declare no conflicts of interest or competing interests.

*Supporting Information.* Phase Diagram of DPPC/DOPC Mixtures in Giant Unilamellar Vesicles with Data; Example calculation for solid area fraction and relative equilibrium surface area; Close up and back-side images of crystals in Figure 1D of main paper; Surface evolver simulation

*Funding*
This work was supported by DOE DE-SC0017870.

*Data Availability* Raw images of each vesicle employed to produce Figure 1D and Figure 2 have been deposited in the Scholarworks database.[55] https://doi.org/10.7275/8121

TOG Graphic

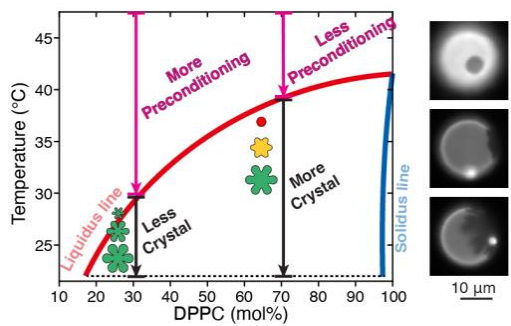

Supporting Information for:

**Sculpting 2D Crystals via Membrane Contractions before and during Solidification**

Hao Wan,[1] Geunwoong Jeon,[2] Gregory M. Grason,[1] Maria M. Santore[1],*

* corresponding author: santore@umass.edu Department of Polymer Science and Engineering, University of Massachusetts 120 Governors Drive Amherst, MA 01003, USA

1. Department of Polymer Science and Engineering, University of Massachusetts
120 Governors Drive Amherst, MA 01003, USA

2. Department of Physics, University of Massachusetts, 710 N. Pleasant St. Amherst, MA 01003, USA

# 1. Phase Diagram of DPPC/DOPC Mixtures in Giant Unilamellar Vesicles with Data

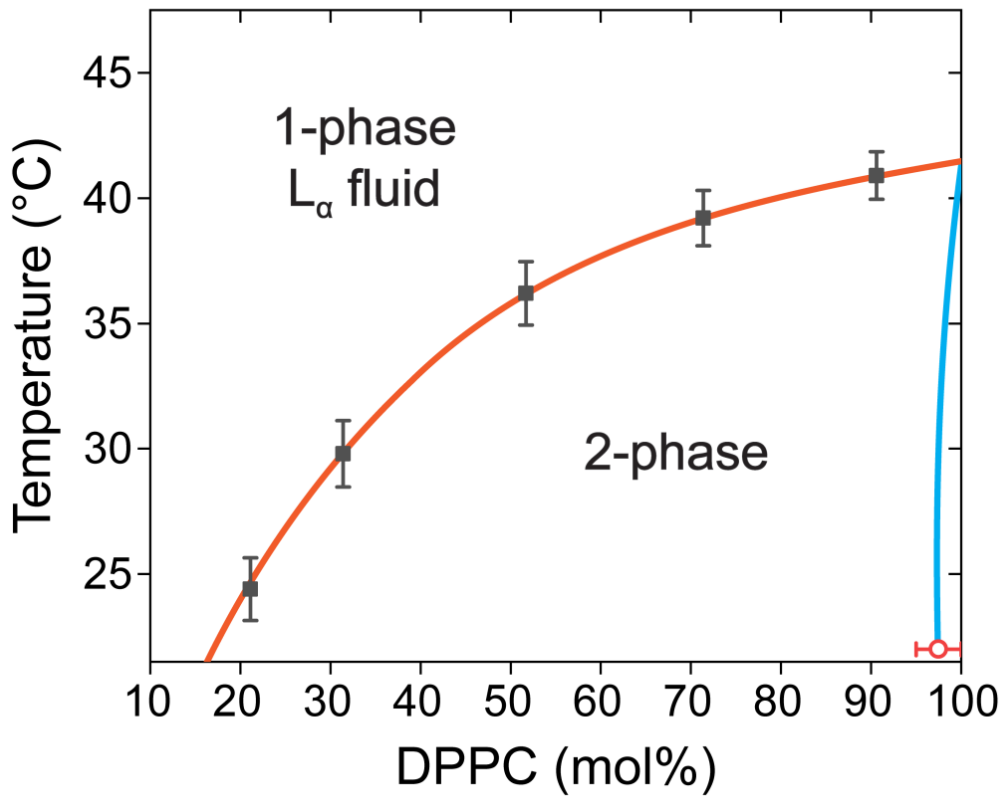

**Figure S1**. Phase diagram for giant unilamellar vesicles of lipid mixtures of DOPC and DPPC and <0.1 mol% tracer. $L_\alpha$ phase boundary (grey squares) based on the first appearance of solid domains upon cooling at each composition. Solid datum (red circle) approximates solid domains known to be nearly pure in DPPC. Error bars are standard deviation and include 6 independent measurement for each composition.

## 2. Example calculation for solid area fraction and relative equilibrium surface area

Here we provide an illustrative example demonstrating the calculation procedure for obtaining the solid area fraction $\phi_s$ and the relative equilibrium area $A_{eq}$ in Figure 1C. This calculation is performed at a specific temperature and DPPC composition, as indicated in Figure 1B.

For instance, at a temperature, T, of 34 °C and an overall DPPC composition ($z_A$) of 50 wt% (mole fraction $z_A = 0.517$, marked by the black point on the illustrative tie line in Figure 1B), we first determine the DPPC composition in the fluid phase ($x_A$). This is achieved using the fitted curve (liquidus line) of $x_A$ as a function of temperature ($x_A = f(T)$), represented by a sixth-order polynomial function, yielding $x_A = 0.432$ at 34 °C( marked by the red point on the example tie line in Figure 1B). The DPPC composition in the solid phase ($y_A$) is approximated as a constant ($y_A = 0.95$), reflecting the known nearly pure DPPC composition in solid domains formed upon phase separation in DOPC/DPPC mixtures.[1, 2]

Employing the inverse lever-arm rule, the solid area fraction $\phi_s$, representing the equilibrium surface area fraction covered by solid domains, can be calculated from Equation (S1):[3]

$$\phi_s = \left(1 + \frac{1}{c}\frac{z_A - y_A}{x_A - z_A}\right)^{-1} \quad (S1)$$

The calculation involves the ratio $c \equiv \underline{A}_S/\underline{A}_L$, which represents the lipid molar area ratio between solid ($\underline{A}_S$) and fluid phases ($\underline{A}_L$). Although this ratio can vary slightly with temperature, it is approximated as a constant here(c = $A_{DPPC}^S$(@ 22 °C)/$A_{DOPC}^F$(@ 22 °C) = 46.4 Å²/69.6 Å²)[3]. This approximation introduces minimal error within the experimental temperature range (22–40 °C). Following these assumptions, at T= 34 °C and $z_A$= 0.517, we calculate a solid area fraction $\phi_s \approx 0.12$.

Next, the relative equilibrium area $A_{eq}$ at a given temperature is computed by numerically solving the ordinary differential equation presented as Equation (2) (also shown as Equation 1B in the main text)

$$\frac{dA_{eq}}{dT} = \left[\kappa_F(1 - \phi_s(T)) + \kappa_s\phi_s(T) - \Delta A^{f-s}\frac{d\phi_s(T)}{dT}\right]A_{eq} \quad (S2)$$

This equation includes the thermal expansion coefficient of the fluid ($\kappa_F$) and solid ($\kappa_s$) phase, the temperature dependent solid area fraction $\phi_s(T)$, and the relative area change during phase transition $\Delta A^{f-s}$. Notably, at a specific DPPC overall composition, $\phi_s$ is defined as a function of temperature based on Equation (1), considering the previously described liquidus line fitting and constant parameters ($z_A$, $y_A$ and c). At temperature above the liquidus line for each composition, $\phi_s$ is set to zero. $\Delta A^{f-s}$ is approximated as 0.17, which is estimated from previous publication,[4] where the change of projected area ratio during phase transition was measured by micropipette aspiration. For practical calculation, the equilibrium area $A_{eq}$ is set as 1 at T = 40 °C. Applying this numerical solution approach at T= 34 °C and $z_A$= 0.517, we obtain $A_{eq} \approx 0.95$.

### 3. Close Up and Back-side Images of Crystals in Figure 1D of Main Paper

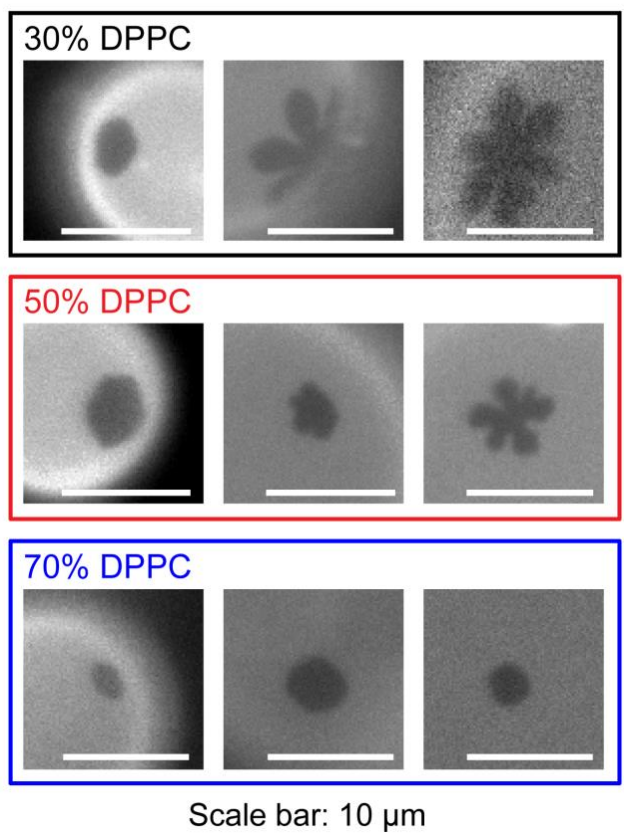

**Figure S2A**. Close-up images of crystals in Figure 1D in main article.

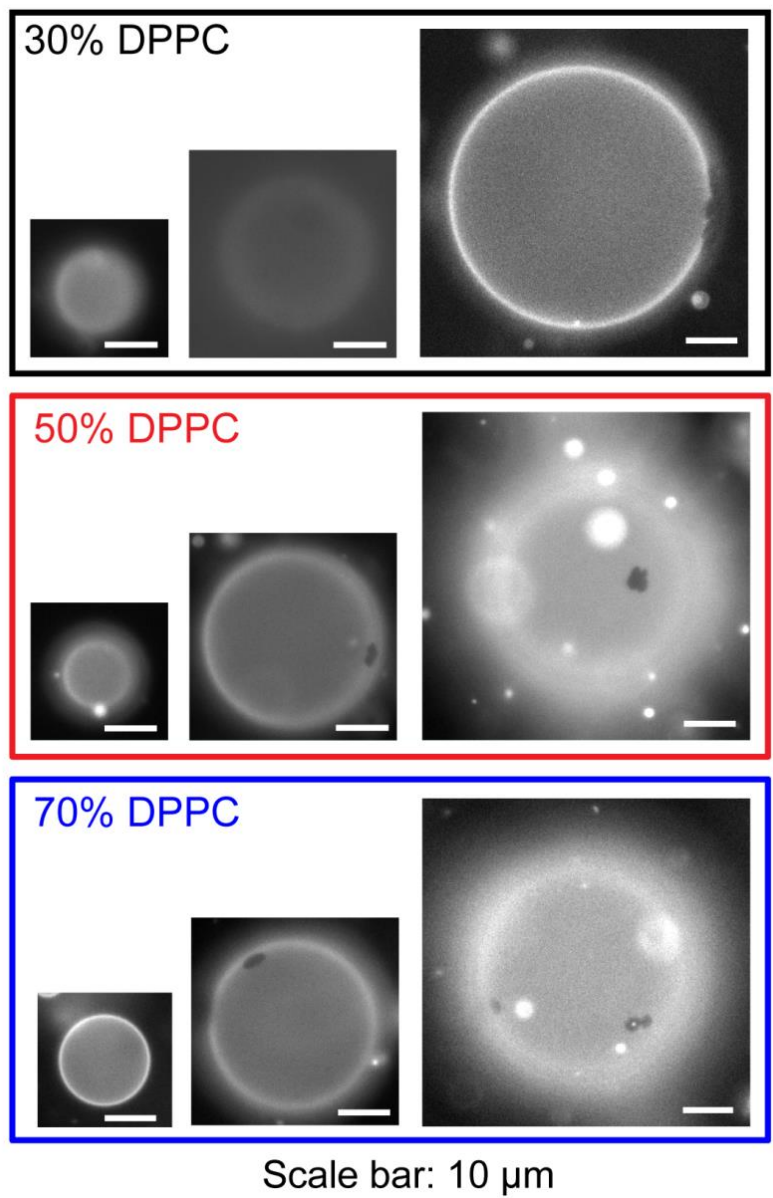

**Figure 2B**. Images of back side (or midplane where relevant) of vesicles from Figure 1D of the main paper.

## 4. Surface Evolver Simulation

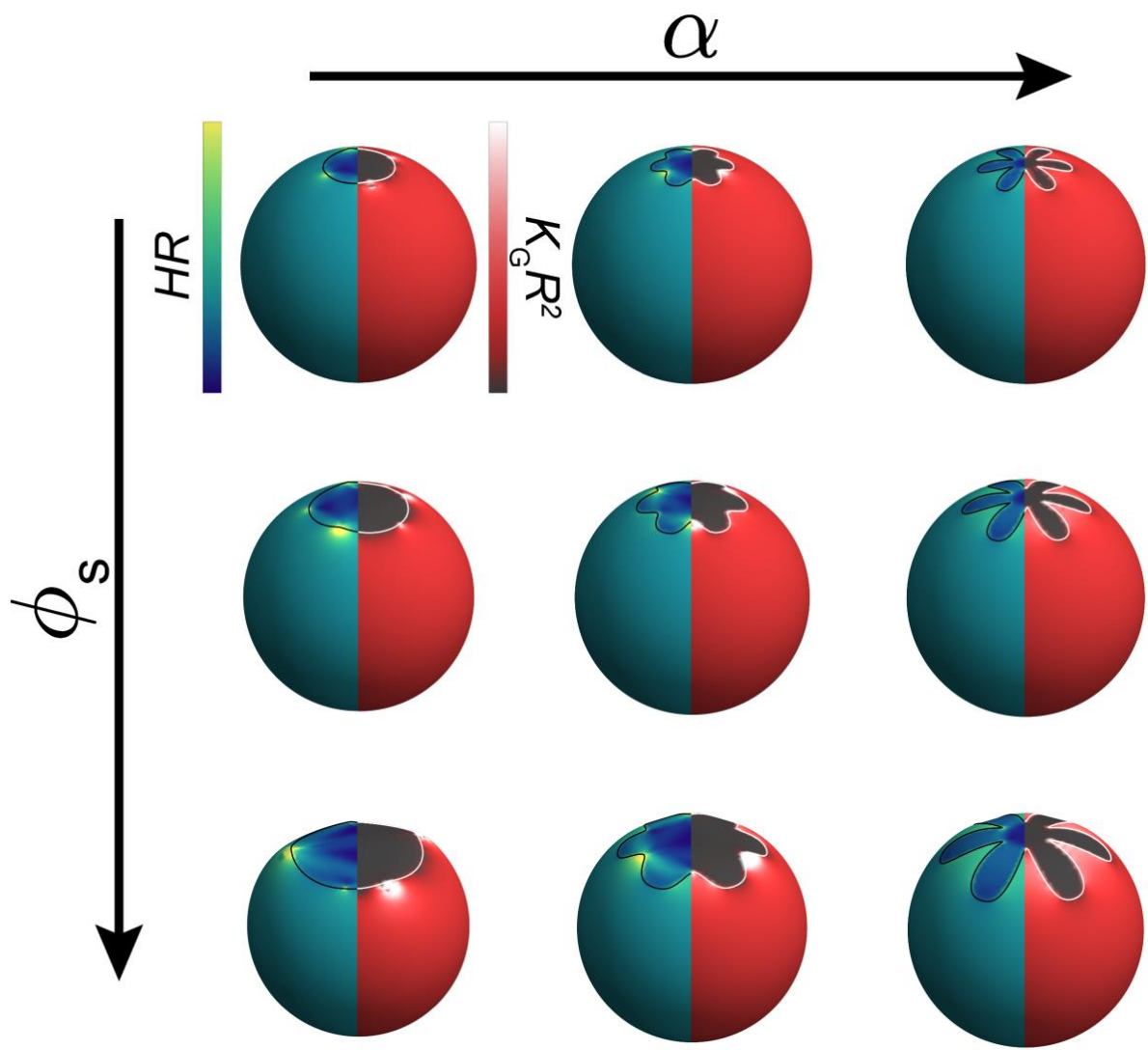

**Figure S3 .** Example structure of equilibrated vesicle models colored by the mean curvature distribution (left) and Gaussian curvature distribution (right). Solid area fraction $\phi_s$ is 0.02, 0.04 or 0.08, increasing to to bottom. Thanks to the small thickness, solid domains strongly expel Gaussian curvature to the fluid membrane leaving nearly vanishing Gaussian curvature over the solid domain.

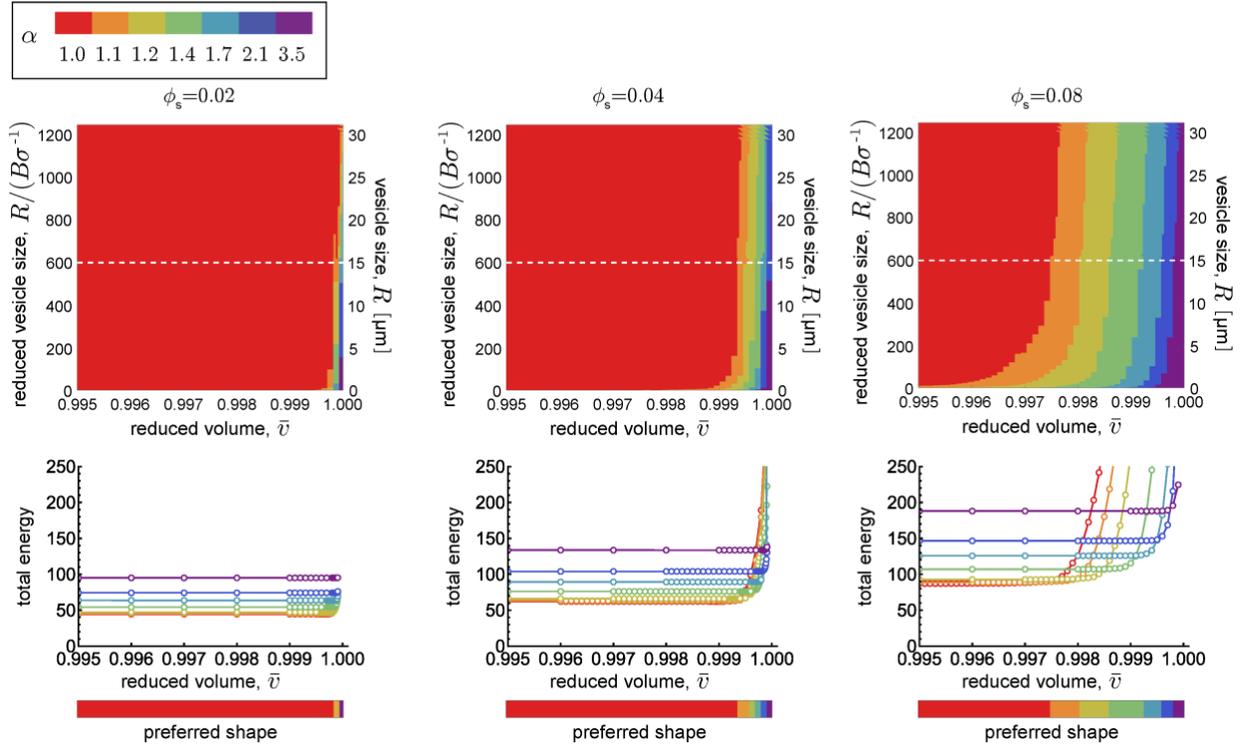

**Figure S4.** Predicted state diagrams (top) of solid domain shapes as functions of vesicle inflation and size, and total reduced energies (bottom) as functions of vesicle inflation for vesicle size corresponding to the white dashed line on the state diagrams. Flower domains become favorable when the elastic energy relaxation from the petaling exceeds the line energy cost $E_{\text{boundary}} = \sigma P$ for the larger solid perimeter. Total energies are in units membrane bending stiffness $B$. $R$ is vesicle radius, establishing the impact of vesicle size.

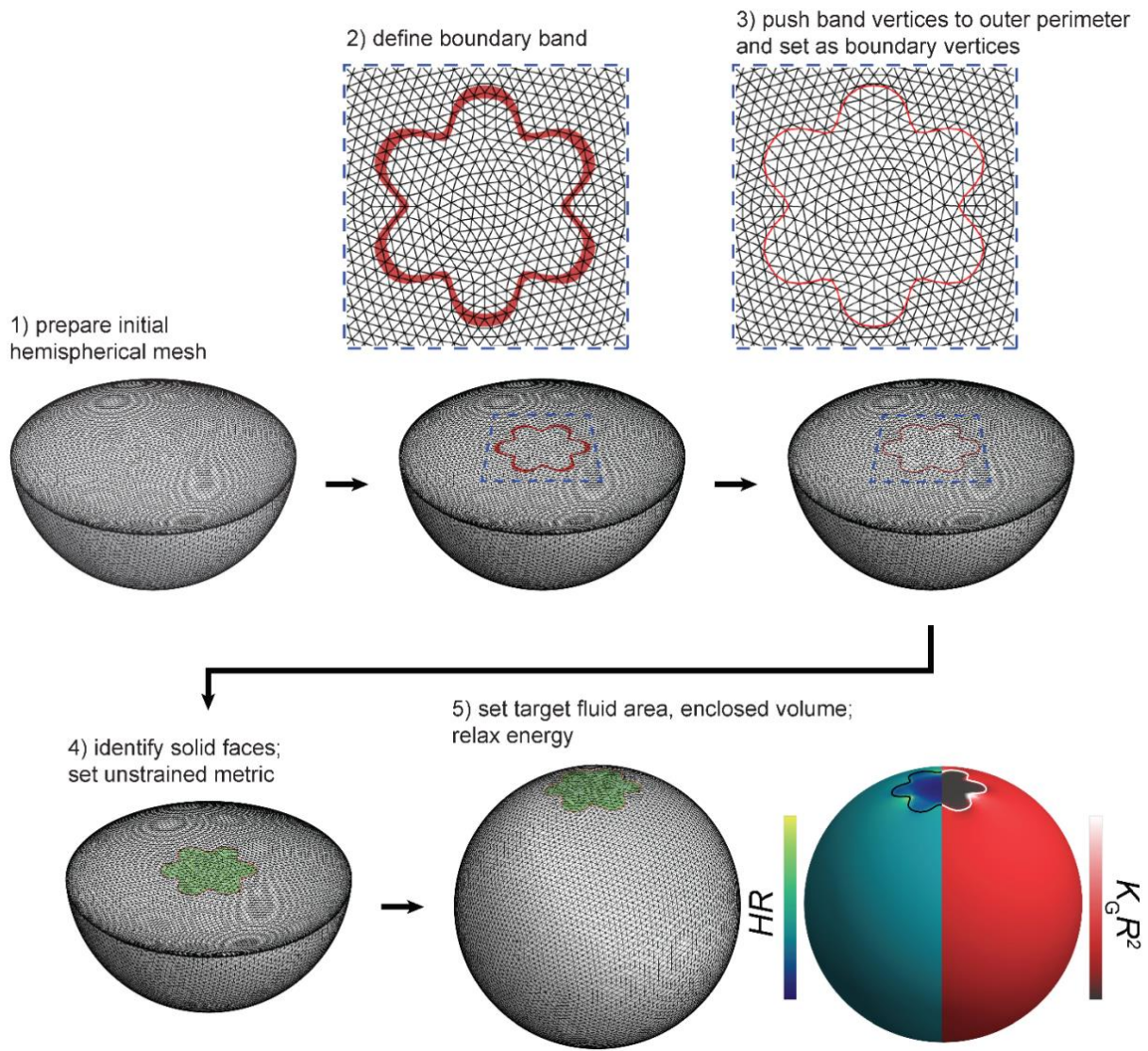

**Figure S5.** Schematics of workflow for Surface Evolver simulations of composite vesicles with flower-like solid domains.

*Simulation protocol.* Here we describe the procedure of numerical simulation of optimal vesicle shapes. The workflow is shown in Figure S5.

1) *Initialize hemispherical mesh* – Prepare initial hemispherical mesh of $\sim 10^4$ triangular facets with the flat top to set the unstrained planar solid domain (i.e. via orthographic projection of the upper hemisphere of a spherical mesh the plane along the equator).
2) *Define the boundary band* – We define a narrow boundary band along the desired solid perimeter. This step is required to match the mesh resolution to small feature sizes of solid domains, particularly at low solid fraction..
3) *Define solid perimeter* – We push the vertices inside of the band to the outer perimeter and set as fluid-solid boundary vertices, and the edges connecting the boundary vertices as boundary edges.
4) *Identify solid faces* – The surface interior to the boundary curve is set as the solid domain, and the exterior is set as the fluid membrane. The unstrained planar solid metric is set at this step, and is used as the reference state to calculate the solid strain energy. Steps to make the triangulation more uniform (e.g. *vertex averaging*) are performed before setting the unstrained metric.
5) *Set target quantities and equilibrate* – Starting from the initial configuration, we set the target fluid area and enclosed volume, and relax the total elastic energy of the entire vesicle.

*Solid domain shape.* The unstrained initial solid domains were modeled as the planar elastic plates with area of $A_{\text{solid}}$ of which perimeters are set by the hexagonally symmetric radial function

$$r(\theta) = r_0 + \frac{1}{2} a r_0 \cos(6\theta) - \frac{1}{10} a r_0 \cos(12\theta) \tag{S3}$$

to mimic the experimentally observed solid domains, where $a$ was varied to set $\alpha = \frac{5+2a}{5-3a}$ values, and $r_0$ adjusts $A_{\text{solid}}$ values. We conducted simulations for seven $\alpha$ values: 1.0 (circular), 1.1, 1.2, 1.4, 1.7, 2.1 and 3.5 (highly petaled).

*Elastic energy.* Following the plate elasticity, the solid domain elastic energy was defined as sum of the strain energy and the bending energy as follows

$$E_{\text{solid}} = \frac{Y}{2(1+\nu)} \int_{\text{solid}} dA \left[ (\text{Tr}\varepsilon)^2 + \frac{\nu}{1-\nu} \text{Tr}\varepsilon^2 \right] + \frac{B}{2} \int_{\text{solid}} dA \, (2H)^2 \tag{S4}$$

where $Y$ is the 2D Young's modulus, $\nu$ is Poisson's ratio, $B$ is the bending modulus, $\varepsilon$ is the 2D strain, and $H$ is the mean curvature. Whereas the fluid membrane elastic energy was defined by the bending energy only as follows

$$E_{\text{fluid}} = \frac{B}{2} \int_{\text{fluid}} dA \, (2H)^2 \tag{S5}$$

imposing lack of the in-plane shear modulus. For the discrete meshes, the elastic energies are approximated as discretized expressions, and attributed to vertices or facets as follows

$$E_{\text{solid}} = Y \sum_j^{\text{facets}^{\text{solid}}} \frac{A_{f,j}}{2(1+\nu)} \left( \text{Tr}[\varepsilon_j^2] + \frac{\nu(\text{Tr}[\varepsilon_j])^2}{1-\nu} \right) + B \sum_i^{\text{vertices}^{\text{solid}}} h_i^2 \frac{A_i}{3} \qquad \text{(S6A)}$$

and

$$E_{\text{fluid}} = B \sum_i^{\text{vertices}^{\text{fluid}}} h_i^2 \frac{A_i}{3} \qquad \text{(S6B)}$$

where subscript $i, j$ denote discretized quantities for vertex $i$ and facets $j$ respectively, and $A_i$ is the effective area for vertex $i$ and $A_{f,j}$ is the area of facet $j$. In *Surface Evolver*, the strain energy was computed by the built-in function *linear_elastic*, and bending energy was computed by the built-in function *star_perp_sq_mean_curvature*. The same bending modulus for solid and fluid was assumed for simplicity. We set the Poisson ratio, $\nu = 0.4$, and the dimensionless thickness, $t/R = \sqrt{B/Y}/R \sim 10^{-4}$ implying relatively much stiffer in-plane strain deformation over out-of-plane bending deformation, where $4\pi R^2$ is defined as the total area of the mesh.

*Energy minimization.* Starting from the initial meshes with unstrained solid domains in Figure S5, we constrained $A_{\text{solid}}$, $A_{\text{fluid}}$, and $V$ to match desired $\phi_s$, and $\bar{v}$. We conducted simulations for three $\phi_s$ values; 0.02, 0.04, 0.08 in the $\bar{v}$ range of 0.98-1.00. The energy relaxation was done repeating the gradient descent steps (via built-in command *g*) and Newton's method (via built-in command *hessian_seek*) until the step size falls below $10^{-9}$.

Note that lack of distinction between local energy minimum and global energy minimum is the nature of the energy minimization method. At the equilibrium state, built-in command *saddle* was applied to check the saddle points, followed by random perturbations (*jiggling*; via built-in command *j*) and *vertex averaging* (via built-in command *V*) to test if the configuration falls into the same local minimum.

**References for Supporting Information**